\documentclass[aps,twocolumn,superscriptaddress,a4paper]{revtex4}

\pdfoutput=1

\usepackage{bm}
\usepackage{graphicx}

\usepackage{color}
\usepackage{amsmath,amssymb,stmaryrd}
\usepackage{siunitx}
\definecolor{mpip} {rgb}{0,0.490196078431,0.478431372549}
\definecolor{mpip2} {rgb}{0,0.690196078431,0.278431372549}
\definecolor{grau} {rgb}{.5,.5,.5}
\definecolor{refokay} {rgb}{.1,.9,.7}

\usepackage{epstopdf}%This line makes .eps figures into .pdf - please comment out if not required.

\definecolor{cream}{RGB}{222,217,201}

\begin{document}

%\begin{widetext}
\parbox{1.985\columnwidth}{
\noindent$\!$\textit{This is a preprint that had been submitted to Soft Matter. The published version can be found under the journal reference} \\[.1cm]
$\textrm{\hspace{1.cm}}$
Soft Matter \textbf{14}, 6809--6821 (2018);\\
$\textrm{\hspace{1.cm}}$
doi: 10.1039/C8SM01051J  \\[.1cm]
\textit{
During the review process, amongst other things, some changes in the references occurred, more details concerning the results of the finite-element simulations were added as a supplementary information, and further references to Refs.~\onlinecite{biller2014modeling} and \onlinecite{biller2015mesoscopic} were added to the text. 
}
}
%\end{widetext}

\vspace{1.cm}

\title{%Magnetically controlled reversible touching and detachment of rigid inclusions in a soft elastic matrix
%Reversible touching and detachment of rigid inclusions in a soft elastic matrix
Reversible magnetomechanical collapse: virtual touching and detachment of rigid inclusions in a soft elastic matrix}

% & \noindent\large{Mate Puljiz,\textit{$^{a}$} Shilin Huang,\textit{$^{b}$}, Karl A. Kalina,\textit{$^{c}$} Johannes Nowak,\textit{$^{d}$} Stefan Odenbach,\textit{$^{d}$} Markus K\"astner,\textit{$^{c}$} G\"unter K. Auernhammer,\textit{$^{b, e}$} and Andreas M. Menzel\textit{$^{a}$}} \\%Author names go here instead of "Full name", etc.

\author{Mate Puljiz}
\affiliation{Institut f\"ur Theoretische Physik II: Weiche Materie, Heinrich-Heine-Universit\"at D\"usseldorf, 40225 D\"usseldorf, Germany.}

\author{Shilin Huang}
\affiliation{Max Planck Institute for Polymer Research, Ackermannweg 10, 55128 Mainz, Germany.}

\author{Karl A. Kalina}
\affiliation{Technische Universit\"at Dresden, Institute of Solid Mechanics, 01062 Dresden, Germany.}

\author{Johannes Nowak}
\affiliation{Technische Universit\"at Dresden, Institute of Fluid Mechanics, 01062 Dresden, Germany.}

\author{Stefan Odenbach}
\affiliation{Technische Universit\"at Dresden, Institute of Fluid Mechanics, 01062 Dresden, Germany.}

\author{Markus K\"astner}
\affiliation{Technische Universit\"at Dresden, Institute of Solid Mechanics, 01062 Dresden, Germany.}

\author{G\"unter K. Auernhammer}
\email{auernhammer@ipfdd.de}
\affiliation{Max Planck Institute for Polymer Research, Ackermannweg 10, 55128 Mainz, Germany.}
\affiliation{Leibniz Institute for Polymer Research, 01069 Dresden, Germany (present address).}

\author{Andreas M. Menzel}
\email{menzel@thphy.uni-duesseldorf.de}
\affiliation{Institut f\"ur Theoretische Physik II: Weiche Materie, Heinrich-Heine-Universit\"at D\"usseldorf, 40225 D\"usseldorf, Germany.}

%\footnotetext{\textit{$^{a}$~Institut f\"ur Theoretische Physik II: Weiche Materie, Heinrich-Heine-Universit\"at D\"usseldorf, 40225 D\"usseldorf, Germany. E-mail: menzel@thphy.uni-duesseldorf.de }}
%\footnotetext{\textit{$^{b}$~Max Planck Institute for Polymer Research, Ackermannweg 10, 55128 Mainz, Germany. }}
%\footnotetext{\textit{$^{c}$~Technische Universit\"at Dresden, Institute of Solid Mechanics, 01062 Dresden, Germany. }}
%\footnotetext{\textit{$^{d}$~Technische Universit\"at Dresden, Institute of Fluid Mechanics, 01062 Dresden, Germany. }}
%\footnotetext{\textit{$^{e}$~Leibniz Institute for Polymer Research, 01069 Dresden, Germany. (Present address). E-mail: auernhammer@ipfdd.de }}

\begin{abstract}
Soft elastic composite materials containing particulate rigid inclusions in a soft elastic matrix are candidates for developing soft actuators or tunable damping devices. The possibility to reversibly drive the rigid inclusions within such a composite together to a close-to-touching state by an external stimulus would offer important benefits. Then, a significant tuning of the mechanical properties could be achieved due to the resulting mechanical hardening. For a long time, it has been argued whether a virtual touching of the embedded magnetic particles with subsequent detachment can actually be observed in real materials, and if so, whether the process is reversible. Here, we present experimental results that demonstrate this phenomenon in reality. Our system consists of two paramagnetic nickel particles embedded at finite initial distance in a soft elastic polymeric gel matrix. Magnetization in an external magnetic field tunes the magnetic attraction between the particles and drives the process. We quantify the scenario by different theoretical tools, i.e., explicit analytical calculations in the framework of linear elasticity theory, a projection onto simplified dipole-spring models, as well as detailed finite-element simulations. From these different approaches, we conclude that in our case the cycle of virtual touching and detachment shows hysteretic behavior due to the mutual magnetization between the paramagnetic particles. Our results are important for the design and construction of reversibly tunable mechanical damping devices. Moreover, our projection on dipole-spring models allows the formal connection of our description to various related systems, e.g., magnetosome filaments in magnetotactic bacteria. 
\end{abstract}

%Please use \dag to cite the ESI in the main text of the article.
%If you article does not have ESI please remove the the \dag symbol from the title and the footnotetext below.
%\footnotetext{\dag~Electronic Supplementary Information (ESI) available: [details of any supplementary information available should be included here]. See DOI: 10.1039/b000000x/}
%additional addresses can be cited as above using the lower-case letters, c, d, e... If all authors are from the same address, no letter is required

%\footnotetext{\ddag~Additional footnotes to the title and authors can be included \emph{e.g.}\ `Present address:' or `These authors contributed equally to this work' as above using the symbols: \ddag, \textsection, and \P. Please place the appropriate symbol next to the author's name and include a \texttt{\textbackslash footnotetext} entry in the the correct place in the list.}

%%%END OF FOOTNOTES%%%

\maketitle

%%%MAIN TEXT%%%%
\section{Introduction}

The fabrication of soft elastic composite materials that consist of rigid particles embedded in a soft elastic environment serves to develop soft actuators \cite{an2003actuating,filipcsei2007magnetic, zimmermann2007deformable,raikher2008shape, fuhrer2009crosslinking,bose2012soft,ilg2013stimuli}, tunable dampers and vibration absorbers \cite{deng2006development,sun2008study,liao2012development, molchanov2014viscoelastic}, components of tunable anisotropic electric conductivity \cite{mietta2016anisotropic}, or devices for energy storage \cite{li2007electric,kim2009high,li2009nanocomposites, wang2011polymer,fredin2013substantial,allahyarov2016dipole}. 
%\GKA{If we make such a wide introduction, I think we can also include old paper on permanent magnetization in soft magnetic gels \cite{collin2003frozen}. }
For instance, actuation is achieved by exposing composites that contain para- or ferromagnetic particles to an external magnetic field gradient \cite{zrinyi1996deformation,collin2003frozen,filipcsei2007magnetic}. 
In this way, forces are directly imposed onto the embedded particles that are drawn into the field gradient \cite{jackson1962classical}. Indirectly, magnetic or electric moments can be induced on the particles by external magnetic or electric fields. 
Then, overall distortions result from the induced mutual particle interactions \cite{diguet2009dipolar,stolbov2011modelling, ivaneyko2012effects,zubarev2013effect,menzel2015tuned, allahyarov2015simulation,huang2016buckling, romeis2016elongated,metsch2016numerical}. 

At the same time, inducing, altering, or reorienting magnetic moments by external magnetic fields affects the overall mechanical properties of the materials due to the modified particle interactions. As a consequence, the static and dynamic elastic moduli are tuned \cite{jolly1996magnetoviscoelastic,jolly1996model, jarkova2003hydrodynamics,filipcsei2007magnetic, stepanov2007effect,bose2009magnetorheological,chertovich2010new, wood2011modeling,ivaneyko2012effects,evans2012highly,han2013field, borin2013tuning,chiba2013wide,pessot2014structural,menzel2015tuned, sorokin2015hysteresis,pessot2016dynamic,volkova2017motion}, and also the nonlinear stress-strain behavior can be qualitatively affected \cite{cremer2015tailoring,cremer2016superelastic}. 

In both situations of actuator applications and tuning the mechanical properties it is often desired to achieve a maximum of the externally induced relative displacements between the particles. 
For inclusions that approach each other during such displacements, the maximum is reached when the particles come into close contact and virtually touch each other. 
%\GKA{Remark: The if we argue that the process is reversible, there must be a polymer layer between the particles of finite thickness, to avoid a singularity in the deformation field.}
In such a situation, the overall material can significantly harden as the steric interactions between the virtually touching particles now come into effect and contribute to overall mechanical stiffness \cite{annunziata2013hardening,pessot2016dynamic}. 
Moreover, as the magnetic interactions strongly increase with reduced distance between the particles, such an approach of the particles can show hysteretic behavior when compared to a subsequent detachment \cite{melenev2006magnetic,stepanov2008motion,biller2014modeling, sorokin2015hysteresis,biller2015mesoscopic,cremer2015tailoring,
cremer2016superelastic,zubarev2017hysteresis,biller2018two}. 
%The implied storage of elastic deformation energy could likewise be interesting for the construction of energy storage devices. 
%\GKA{Sorry, I don't get the last point. I this meant to be a political remark, or do we have a concrete model system in mind? }

Experimentally, observing and tracking externally induced rearrangements of magnetic particles within elastic environments is standard. 
For example, this concerns active magnetic microrheology \cite{crick1950physical,ziemann1994local,bausch1999measurement, waigh2005microrheology,wilhelm2008out, chippada2009complete,roeder2012shear,bender2013determination, huang2017structure}, or, in fact, quantifying the structural rearrangement within samples of magnetic elastic composite materials \cite{an2014conformational,gundermann2014investigation, huang2016buckling,gundermann2017statistical}. Recently, in different samples of initially rather isotropically distributed and well separated inclusions, the formation of short chain-like arrangements has been observed experimentally \cite{stepanov2007effect,stepanov2008motion,gundermann2017statistical,schumann2017characterization}. Still, in the final state of the induced textures in Ref.~\citenum{gundermann2017statistical}, there are gaps between the individual particles. 
If the particles could be driven basically into contact by further increasing the magnetic interactions between them, %simplified theoretical minimal models %, more details of which are given below, 
%predict 
a strong increase of the elastic moduli may be observed as indicated above. % \cite{}. %In this situation, the harder steric interactions would come into play. 

To really observe such an approach in actual experiments has so far been difficult to conceive. On the one hand, elastic polymeric systems are typically rather incompressible. Yet, the material between the particles would need to be significantly compressed or be squeezed out laterally. On the other hand, the elastic matrix could be ruptured by the particle approach to allow a touching of the particles. However, in this case, the desired reversibility of the whole process, to allow repeated usage in practice, is questioned. 

In spite of these legitimate concerns, we have found that the described process is possible and actually observable in experiments. We have embedded two magnetizable nickel particles into a soft elastic polymeric gel matrix. Stepwise increasing the magnetic attraction between the two particles by an external magnetic field, the particles approached until they finally snapped together into close contact. Switching off the magnetic attraction, the particles returned to their initial state. The process appeared reversible, and we were able to start this cycle repeatedly from the beginning.  

Reduced minimal models that come into question to describe such situations address rigid spherical magnetic particles of finite size.  %that magnetically attract and repel each other. At the same time, 
In such models, some elastic contribution, representing the embedding elastic environment, tries to maintain a preset distance between the two particles. This can be achieved by an elastic spring-like interaction \cite{stepanov2008motion,annunziata2013hardening,pessot2014structural,tarama2014tunable,ivaneyko2015dynamic, allahyarov2015simulation,allahyarov2016dipole,pessot2016dynamic, cremer2017density,pessot2018tunable}, more refined contributions such as elastic bars and rods  \cite{biller2014modeling,biller2015mesoscopic}, or discretized volume elements describing the elastic matrix \cite{cremer2015tailoring,cremer2016superelastic, kalina2016microscale,metsch2016numerical,attaran2017modeling}. 
Apart from that, the elastic restoring forces resulting from deformations of the surrounding elastic environment can also be calculated analytically in the framework of linear elasticity theory \cite{phan1993rigid,kim1994faxen,phan1994load,norris2008faxen, puljiz2016forces,puljiz2017forces,menzel2017force}. 
Often, the magnetic moments on the particles are represented by %permanent and constant
magnetic dipoles of permanent magnitude \cite{klapp2005dipolar,holm2005structure,diguet2009dipolar, wood2011modeling,han2013field,annunziata2013hardening, pessot2014structural,menzel2014bridging,tarama2014tunable,ivaneyko2015dynamic, pessot2015towards,pessot2016dynamic,cremer2015tailoring, cremer2016superelastic}. Addressing magnetization of dipolar particles by a non-saturating external magnetic field is possible by including ``loop corrections'', i.e., an iterative numerical loop that calculates the additional contributions to the dipole moments resulting from the mutually induced magnetization between the magnetic particles \cite{zubarev2013magnetodeformation,allahyarov2016dipole}. More refined approaches resolve the finite size of the particles and take into account spatial variations of the induced magnetization across the particle interior \cite{han2013field,biller2015mesoscopic,kalina2016microscale, metsch2016numerical,romeis2017theoretical}. 

Below, we first report our experimental observations in Sec.~\ref{experimental}. We then continue by an analytical description of the situation in terms of linear elasticity theory in Sec.~\ref{linear-elastic}. Despite the strong relative distortions, we find that the experimental results can be described reasonably well by the linearly elastic approach. An effective local elastic modulus for the elastic matrix can be extracted in this way. This allows to map the whole situation to simplified dipole-spring approaches in Sec.~\ref{dipole-spring}. Moreover, we predict significant hysteretic behavior on this basis. After that, we present results from significantly 
more detailed finite-element (FE) simulations in Sec.~\ref{xfem}. They include nonlinear contributions to the response of the elastic matrix and resolve the magnetization across the inside of the magnetic particles. Particularly, they support the analytical approaches in the still separated state, but imply quantitative corrections close to touching of the particles and for the predicted hysteretic behavior. %In Sec.~\ref{hysteresis}, using the different methods, we discuss and predict hysteretic behavior. %, which so far could not be resolved experimentally. 
Several conclusions are given in Sec.~\ref{conclusion}.

\section{Experimental observations}
\label{experimental}

\begin{figure}
\centerline{\includegraphics[width=\columnwidth]{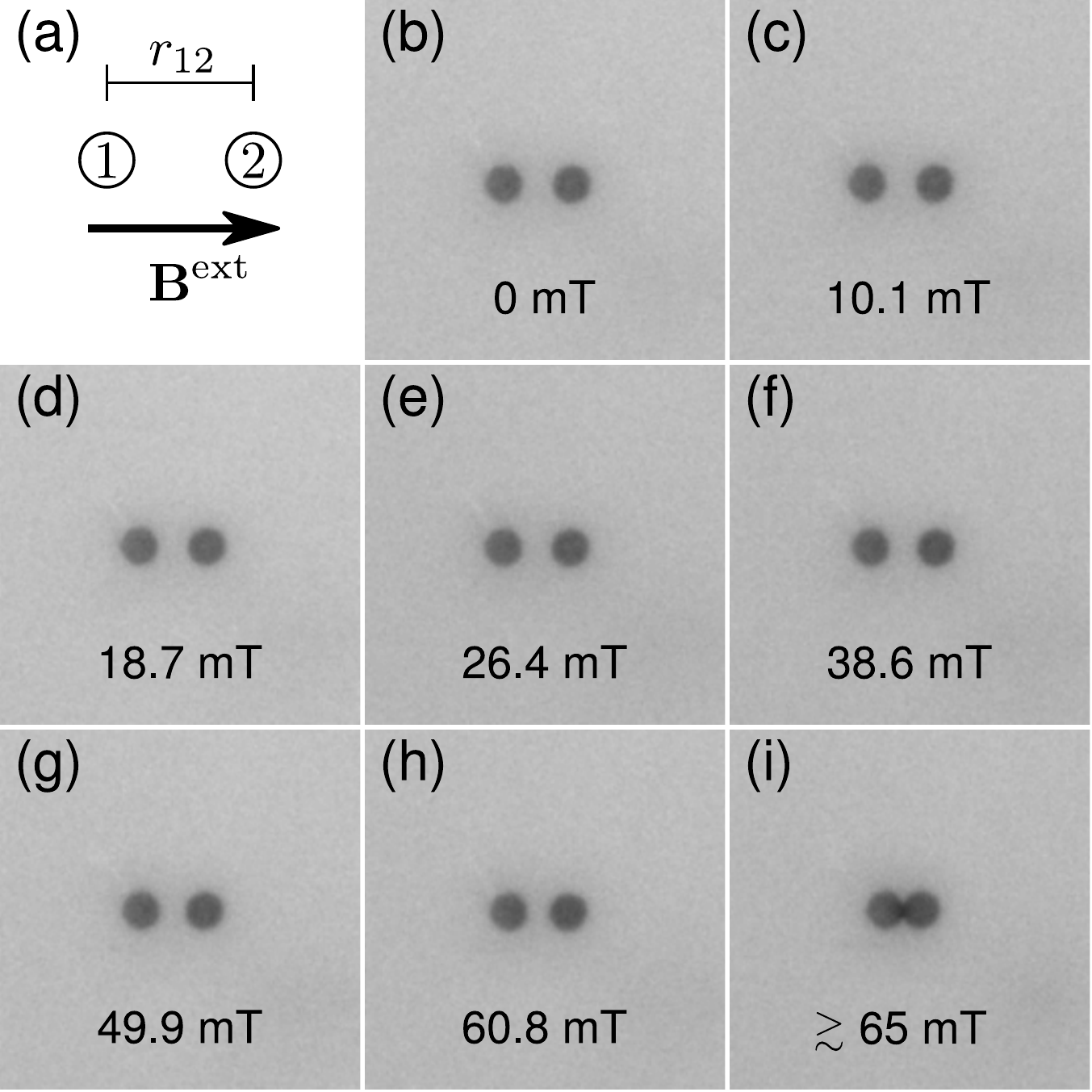}}
\caption{(a) Schematic illustration of our set-up. Two magnetic nickel particles are embedded in an elastic polymer matrix, separated by a distance $r_{12}$. An external magnetic field $\mathbf{B}^{\text{ext}}$ is applied along the center-to-center vector of the two particles. (b--i) Experimental snapshots for increasing the magnetic field strength as indicated by the given magnitudes. At high enough field strengths (i) the particles come into close contact. This snapping together is reversible, that is, switching off the field, the particles separate again (b). 
}
\label{setup}
\end{figure}

Our experimental system was generated in a way similar to the one presented in Ref.~\citenum{puljiz2016forces}. Two nickel particles (purchased from Alfa Aesar $ - 100 \pm 325$ mesh, purity 99.8\%) were embedded in the center plane of a polydimethylsiloxane-based soft gel \cite{huang2016buckling} enclosed in a plastic mold (diameter $\sim \SI{24}{\milli\metre}$, height $\sim \SI{6}{\milli\metre}$). %To place the particles in a well-defined plane in the middle of the soft gel 
For this purpose, we first filled the sample volume only to half height with the reaction solution and allowed for a first cross-linking (for about $\SI{0.5}{\hour}$). We then placed the particles by hand on top of the surface of this first gel layer with a center-to-center distance of initially $r_{12} = (302.4\pm{1.9})\:\SI{}{\micro\metre}$ as defined in Fig.~\ref{setup}~(a). 
Thereafter, we filled in the top half layer of the gel. To guarantee a good connection between the two gel layers, a cross-linking for at least $7$ days at room temperature was allowed. Despite a careful adjustment of the composition, the final modulus had some uncertainty and could not be measured by a rheometer on a separate sample. The underlying unavoidable variations in the composition resulted from the close proximity to the percolation threshold of cross-linking. 

\begin{figure}
\centerline{\includegraphics[width=\columnwidth]{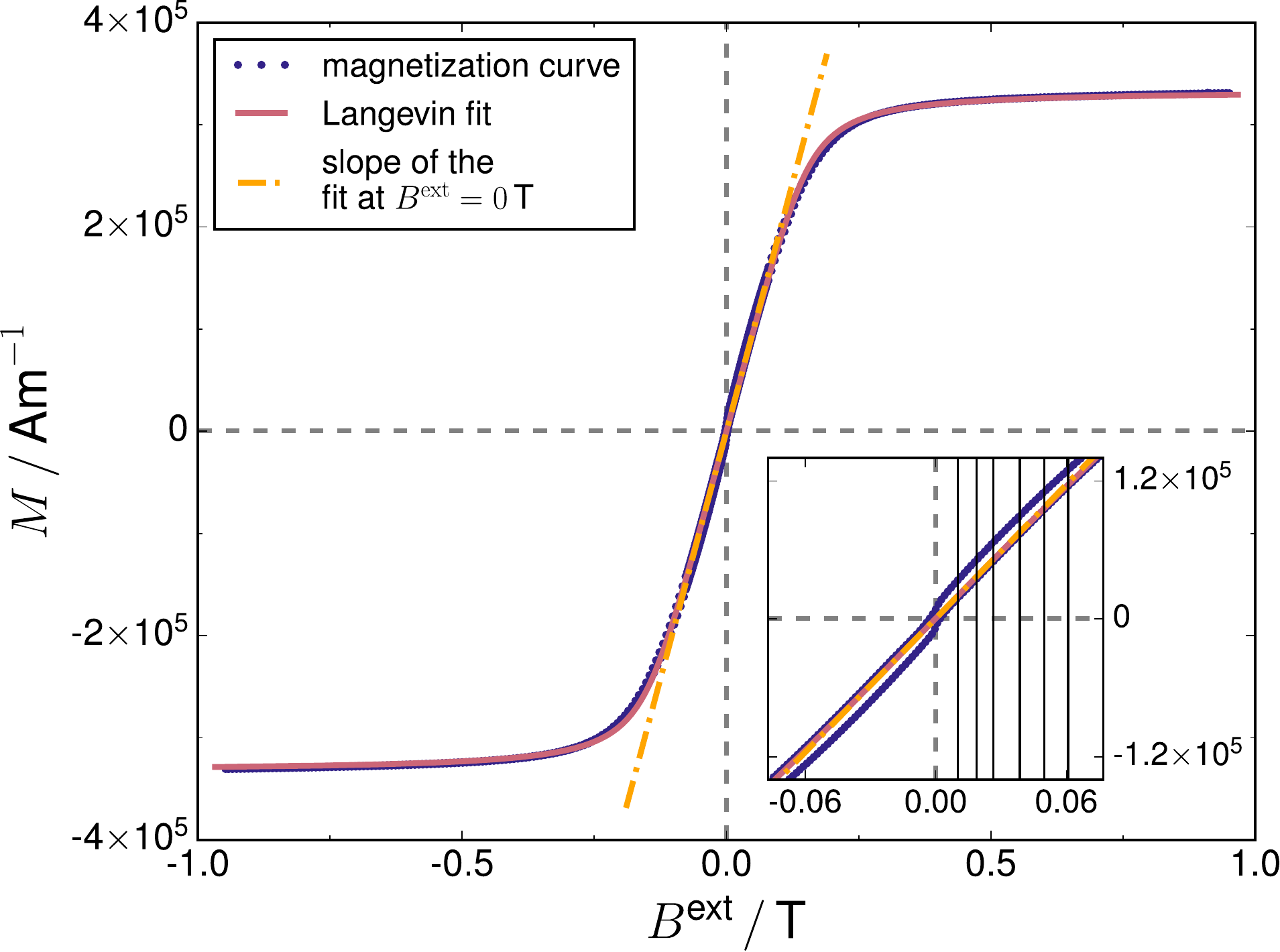}}
\caption{Bulk magnetization curve $M(B^\text{ext})$ for the investigated nickel particles. The experimentally measured data curve (blue dots) reveals minor remnant magnetization, where a small hysteresis between increasing and decreasing external magnetic field can be identified (see also the inset). % (the more isolated blue dots represent the first magnetization run \amm{stimmt das???} \GKA{\em Sorry, I don't get the remark in this brakets}). 
A fit of the data points using a Langevin function, see Eq.~(\ref{eq:Langevin}), is performed (red line).
Linearizing the function, a slope of the curve at $B^\text{ext}=\SI{0}{T}$ is found as indicated by the orange dash-dotted line.
It leads to a value for the relative permeability of $\mu_r= 14.10\pm0.58$. 
In the inset, the applied field strengths in our two-particle experiment are marked by the vertical lines.}
\label{fig2}
\end{figure}

Our nickel particles had a diameter of $170 \pm \SI{10}{\micro\metre}$. Fig.~\ref{fig2} shows the corresponding magnetization curve determined by a vibrating sample magnetometer, Lake Shore 7407. 
It indicates a small remanence of $\sim \SI{7.5}{\kilo\ampere/\metre}$ and a low  coercive field of $\sim \SI{2.4}{\milli\tesla}$. In our experiment, an external magnetic field was applied along the center-to-center vector of the two particles, see Fig.~\ref{setup}~(a). This field was generated by a Halbach array of four magnets. To allow for an adjustable magnitude of the field, the magnets in the array could be moved radially in discrete steps from distances of $\sim \SI{2}{\centi\metre}$ to  $\sim \SI{10}{\centi\metre}$ from the center of the sample. In this way, the magnetic field strength could be varied in a range from $\SI{0}{\milli\tesla}$ to $\sim\SI{100}{\milli\tesla}$. For each used position of the magnets, the magnetic field at the sample center was measured by a Lake Shore Model 425 Gaussmeter with a transverse probe \cite{huang2016buckling}. %The field orientation was chosen to be parallel to the center-to-center vector to the two particles. 

In the experiment, the magnetic field was increased in 6 steps from $\SI{0}{\milli\tesla}$ to $\SI{60.8}{\milli\tesla}$. After applying the next possible field strength $\sim \SI{65.1}{\milli\tesla}$, the particles reversibly snapped into close contact. To determine the center-to-center distance of the particles, their positions were imaged using a CCD camera (MATRIX VISION mvBlueCOUGAR-S) equipped with a zoom macro lens (Navitar Zoom 7000). The particle tracker module of ImageJ was employed \cite{imagej}. Due to little variations of the image quality, the uncertainty in the measured particle positions varied slightly during the experiment, but was always below 4\% of the particle diameter. 

\begin{figure}
\centerline{\includegraphics[width=\columnwidth]{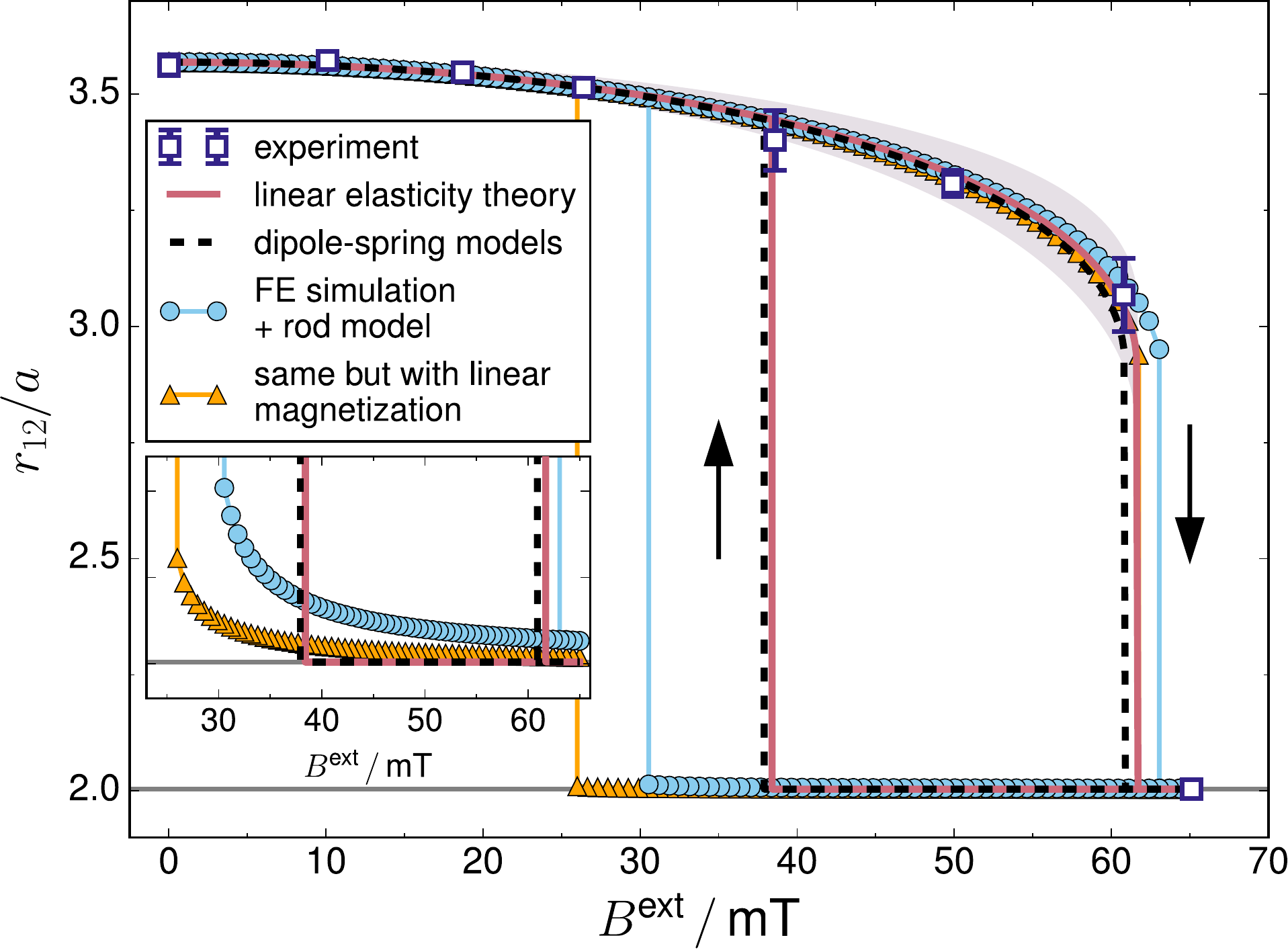}}
\caption{Interparticle center-to-center distance $r_{12}$ as a function of the magnitude $B^\text{ext}$ of the externally applied magnetic field.
Open squares with error bars mark the experimentally measured distances as a function of the applied magnetic field.
The solid (red) curve is obtained analytically from linear elasticity theory  and a magnetic dipole model as described in Sec.~\ref{linear-elastic}. Fitting to the experimental data points, we extract a shear modulus of $\mu=226.0\pm2.8$~Pa. The shaded area corresponds to the uncertainty in the theoretical result because of the uncertainties in the fit and the experimental input parameters \cite{puljiz2016forces}. A collapse of the separated state is predicted at $B^\text{ext}\approx \SI{61.7}{\milli\tesla}$ ($r_{12}/a=2$ corresponds to a touching state where the particle surfaces are in contact). 
For decreasing magnetic field strength, the theory indicates pronounced hysteresis and a detachment at $B^\text{ext}\approx\SI{38.3}{\milli\tesla}$.
We have mapped the theory to simplified dipole-spring models as described in Sec.~\ref{dipole-spring}, represented by the dashed line. The curve shows good agreement with the results obtained from linear elasticity theory.
Apart from that, we performed additional finite-element simulations as detailed in Sec.~\ref{xfem} and marked by the light blue data points.
They include effects of nonlinear elasticity of the polymeric matrix, nonlinear magnetization, and spatially resolved magnetization within the spherical particles.
To good approximation, they confirm the field magnitude at which the particles approach. As may have been expected, they reveal an even more pronounced hysteresis relatively to the linearly elastic theory. %by comparing the field magnitude of particle separation. 
Additionally, we have included simulation results for linearized magnetization behavior as marked by the (orange) triangles. 
Moreover, as shown by the inset, the simulations indicate that a finite gap may remain between the particles in the collapsed state.
}
\label{fig3}
\end{figure}

As a result, we observed that the center-to-center distance with increasing magnetic field smoothly decreased up to a field strength $\sim \SI{60.8}{\milli\tesla}$. Fig.~\ref{fig3} contains the measured experimental data points and Fig.~\ref{setup}~(b--i) shows the corresponding snapshots. 
Applying fields of strength $\sim \SI{65.1}{\milli\tesla}$, we observed that the particles snap into close contact, see Fig.~\ref{setup}~(i). However, this approach is reversible. Switching off the external magnetic field, the particles separated again and took their initial position, at least within the experimental errors. We also repeated the procedure several times, reversibly observing the snapping at the same magnetic field strength. 

An interesting question is whether there is some hysteresis involved in the approach of the particles. Does the separation of the particles, when continuously reducing the strength of the magnetic field, occur at lower field magnitudes than when increasing the magnetic field? 

Unfortunately, our experimental set-up does not allow to clarify this question. To alter the field strength, we have to take away the magnets, readjust their holders, and then reinsert the magnets at the new distance from the center of the sample. %Therefore, we cannot maintain the magnetic field acting on the sample while readjusting its strength. 
The reinsertion always corresponds to an increase of the magnetic field amplitude, with an intermediate state of vanishing field. It is therefore up to theoretical approaches and simulations to clarify the question about possible hysteresis when decreasing the magnetic field.

\section{Description by linear elasticity theory}
\label{linear-elastic}

Consequently, we now compare our experimental results in Fig.~\ref{fig3} with different theoretical approaches. In this way, we can extract from the theories the effective local elastic shear modulus $\mu$ that, as described above, could not be measured by a rheometer. Using this result, we will be able to address the question of underlying hysteresis. % in Sec.~\ref{hysteresis}, which so far could not be resolved by the available experimental apparatus. 

We start by a description in the framework of linear elasticity theory \cite{landau1986theory}. The advantage of this approach is that, assuming a homogeneous, isotropic, and infinitely extended elastic matrix, the situation of two displaced spheres can, in principle, be solved analytically to any desired accuracy. On the downside, of course, only a linearly elastic response of the matrix is described. In Fig.~\ref{fig3}, the experimental data points before the collapse are confined to relative distance changes $\lesssim 15$~\% between the particle centers, in favor of a linearly elastic characterization of the investigated soft elastic gel matrix. 
Moreover, we consider the elastic matrix to be incompressible and set the Poisson ratio $\nu=0.5$. This, likewise, represents a reasonable approximation for the system under investigation. 

Our two idealized rigid spherical inclusions, both of radius $a$ and labeled by $1$ and $2$, are initially centered at positions $\mathbf{r}_1$ and $\mathbf{r}_2$, respectively.
If an external magnetic field $\mathbf{H}^\text{ext}$ is applied from outside, the spheres are magnetized. 
To present our results, we use the quantity $\mathbf{B}^\text{ext}=\mu_0\mathbf{H}^\text{ext}$, where $\mu_0$ is the magnetic vacuum permeability. 
For an infinitely extended bulk material, the corresponding magnetization curve is given by Fig.~\ref{fig2}.

%\karl{%
Assuming magnetic isotropy within our particles, we describe their internal magnetization $\mathbf M$ by the Langevin function 
\begin{equation}
	\mathbf M = M^\text{s}\Bigg[\coth\big(\alpha |\mathbf H^\text{in}|\big)-\frac{1}{\alpha |\mathbf H^\text{in}|}\Bigg] \frac{\mathbf H^\text{in}}{|\mathbf H^\text{in}|}, \label{eq:Langevin}
\end{equation}
where $M^\text{s}$, $\alpha$, and $\mathbf{H}^\text{in}$ denote the magnitude of the saturation magnetization, a scaling parameter, and the magnetic field inside the material, respectively.
Considering only one spherical particle exposed to a homogeneous external magnetic field $\mathbf{H}^\text{ext}$, also $\mathbf{M}$ and $\mathbf{H}^\text{in}$ are homogeneous. Because of the spherical geometry, a demagnetization effect with a demagnetization factor of $1/3$ occurs, so that \cite{jackson1962classical}
\begin{equation}\label{H_in0}
	\mathbf{H}^\text{in} ={} \mathbf{H}^\text{ext}-\frac{1}{3}\mathbf{M}.
\end{equation}
Fitting Eq.~(\ref{eq:Langevin}) to the experimental data given in Fig.~\ref{fig2}, we find the values \mbox{$M^\text{s}=(3.333 \pm 0.290)\times 10^5 \:\SI{}{\ampere\metre}^{-1}$} and \mbox{$\alpha=(1.179 \pm 0.057)\times10^{-4}\:\SI{}{\metre\ampere}^{-1}$}, respectively.
%}%

The experimentally measured data points displayed in Fig.~\ref{fig3} before the collapse were obtained at $B^\text{ext}\lesssim \SI{61}{\milli\tesla}$, and the collapse of the particles occurred below $\SI{65.1}{\milli\tesla}$. Fig.~\ref{fig2} demonstrates that the overall magnetization curve of the material in this regime can be well approximated by a straight line. 
We may thus treat the problem within the framework of linear magnetization behavior by \cite{jackson1962classical}%
\begin{equation}\label{magnetization_1}
	\mathbf{M} ={} (\mu_r-1)\mathbf{H}^\text{in},
\end{equation}
with $\mu_r$ the relative magnetic permeability \cite{jackson1962classical}.
A linearization of Eq.~(\ref{eq:Langevin}) for small values of $H^\text{in}$ leads to $\mu_r = \frac{\alpha M^\text{s}}{3}+1$, so that in our case $\mu_r=14.10\pm0.58$.

Combining Eqs.~(\ref{H_in0}) and (\ref{magnetization_1}) yields
\begin{equation}\label{HM}
	\mathbf{H}^\text{in} = \frac{3}{\mu_r+2}\mathbf{H}^\text{ext},
\qquad
	\mathbf{M} = 3\frac{\mu_r-1}{\mu_r+2}\mathbf{H}^\text{ext}.
\end{equation}
Outside the spherical particle, the induced magnetic field resulting from the internal magnetization coincides with the one of a magnetic point dipole \cite{jackson1962classical}
\begin{equation}\label{eq_m}
\mathbf{m}_i=\frac{4\pi}{3} a^3\mathbf{M},
\end{equation} 
located at the center of the sphere. 

Since the magnetic particles in the states corresponding to the experimental data points before the collapse are well separated from each other, we maintain this picture also in our situation of two particles.  That is, we approximate the magnetic moment of each particle by a point dipole $\mathbf{m}_i$, $i=1,2$, located at the particle center. 
%However, for weak external fields, as is the case here, the magnetization is not saturated, i.e., the dipole moment of each particle changes under the magnetic field created by the induced dipole of the respective other sphere. Therefore, an iterative scheme has to be introduced in order to find the resulting final dipole moments $\mathbf{m}_i$.
%As stated above, in our two-particle system, the magnetic moment $\mathbf{m}_i$ induced in particle $i$ not only depends on the external field 
%
However, the role of $\mathbf{H}^\text{ext}$ in Eq.~(\ref{HM}) is now not only played by the external magnetic field itself. Also the field induced by the other dipole contributes at this point. We denote the magnetic field induced by dipole $\mathbf{m}_j$ at position $\mathbf{r}$ as 
\begin{equation}\label{H_dipole}
	\mathbf{H}^\text{dip}_j(\mathbf{r}) = \frac{1}{4\pi} \left(\frac{3(\mathbf{r}-\mathbf{r}_j)\; \mathbf{m}_j\cdot(\mathbf{r}-\mathbf{r}_j) }{|\mathbf{r}-\mathbf{r}_j|^5}-\frac{\mathbf{m}_j}{|\mathbf{r}-\mathbf{r}_j|^3}\right).
\end{equation}
Together, from Eqs.~(\ref{HM})--(\ref{H_dipole}), we obtain
\begin{equation}\label{magnetic_moment}
	\mathbf{m}_i ={} 4\pi a^3 \frac{\mu_r-1}{\mu_r+2}\left[\mathbf{H}^\text{ext} + \mathbf{H}^\text{dip}_j(\mathbf{r}_i) \right].
\end{equation}
Obviously, Eqs.~(\ref{H_dipole}) and (\ref{magnetic_moment}) need to be solved by iteration \cite{zubarev2013magnetodeformation,allahyarov2016dipole}. 
%In the first step of iteration, neither sphere is magnetized and $\mathbf{H}_j^\text{dip}$ in Eq.~(\ref{magnetic_moment}) equals zero.In the next step, due to the non-zero external field, dipolar fields as given by Eq.~(\ref{H_dipole}) are induced, thereby changing the overall field. Thus, Eq.~(\ref{H_dipole}) has to be reinserted into Eq.~(\ref{magnetic_moment}), yielding different magnetic moments than in the first step.
After convergence, the final magnetic moments are obtained for a certain particle configuration. 
For the dipole configurations resulting for the experimental data points in Fig.~\ref{fig3} we have estimated $H^\text{dip}/H^\text{ext}\lesssim 6$~\%, which supports our approximation in terms of magnetic dipoles. 

%, from which the dipole-dipole interaction forces from Eq.~(\ref{magnetic_force}) can be calculated.
%Therefore, magnetic dipole-dipole interaction forces act onto the particles and drive them out of their equilibrium positions.
%These forces have the form \cite{jackson1962classical}
%
%\begin{eqnarray}\label{magnetic_force}
%	\mathbf{F}_i &={}& -\frac{3\mu_0}{4\pi r^4_{ij}} \Big[5 \mathbf{\hat{r}}_{ij} (\mathbf{m}_i\cdot\mathbf{\hat{r}}_{ij})(\mathbf{m}_j\cdot\mathbf{\hat{r}}_{ij}) - \mathbf{\hat{r}}_{ij}(\mathbf{m}_1\cdot\mathbf{m}_j)\notag\\
%	&{}&- \mathbf{m}_i (\mathbf{m}_j\cdot\mathbf{\hat{r}}_{ij})- \mathbf{m}_j (\mathbf{m}_i\cdot\mathbf{\hat{r}}_{ij})\Big],
%\end{eqnarray} 
%
%with $\mathbf{r}_{ij}=\mathbf{r}_i-\mathbf{r}_j$, $r_{ij}=|\mathbf{r}_{ij}|$, $\mathbf{\hat{r}}_{ij}=\mathbf{r}_{ij}/r_{ij}$, and $i,j\in\{1,2\,|\,i\neq j\}$.

Next, we need to determine the magnetic forces resulting from the mutual interaction between the two induced magnetic moments. For our set-up and in the regime of linear magnetization, a corresponding expression for the change in overall magnetic energy when including the magnetic particles into the external magnetic field \cite{jackson1962classical} reads 
\begin{equation}
\label{Wmag}
W^{\text{mag}} = {}-\frac{1}{2}(\mathbf{m}_1+\mathbf{m}_2)\cdot\mathbf{B}^{\text{ext}}. 
\end{equation}
Here, $\mathbf{m}_1$ and $\mathbf{m}_2$ are the magnetic moments obtained from the above iteration and contain the mutual magnetization between the two particles. The magnetic forces on the two particles then are obtained as 
\begin{equation}
\mathbf{F}_i = -\nabla_{\mathbf{r}_i}W^{\text{mag}}
\label{FnablaW}
\end{equation}
for $i=1,2$. Our set-up is axially symmetric with respect to the center-to-center direction $\mathbf{\hat{r}}_{ij}=\mathbf{r}_{ij}/r_{ij}$, with $\mathbf{r}_{ij}=\mathbf{r}_i-\mathbf{r}_j$ and $r_{ij}=|\mathbf{r}_{ij}|$. Moreover, it is symmetric with respect to the center position between the two particles that we choose as the origin. Then, we may discretize Eq.~(\ref{FnablaW}) as
\begin{equation}
\mathbf{F}_i = -\mathbf{\hat{r}}_{i}
\lim\limits_{\Delta\searrow 0} 
\frac
	{W^{\text{mag}}|_{(r_i+\Delta)\mathbf{\hat{r}}_i} 
	- W^{\text{mag}}|_{(r_i-\Delta)\mathbf{\hat{r}}_i}
	}
	{2\Delta},
\end{equation}
where $i\in\{1,2\}$, $\mathbf{r}_i=r_i\mathbf{\hat{r}}_i$ with $r_i=|\mathbf{r}_{i}|$, and $\mathbf{\hat{r}}_{1}=-\mathbf{\hat{r}}_{2}=\mathbf{\hat{r}}_{12}$. In this expression, the $\mathbf{m}_i$ for each modification in position $\Delta$ need to be reevaluated by the iterative procedure described above. 

As the magnetic forces act on the particulate inclusions, the particles are pressed against the surrounding elastic matrix. This leads to matrix deformations and, in turn, to elastic restoring forces on the particles that limit their induced displacements. 
Moreover, the final displacements $\mathbf{U}_i$ of the particles are elastically coupled to each other through the induced matrix deformations. 
If one of the particles distorts the environment, the other particle, likewise embedded in the elastic matrix, is displaced together with the induced matrix relocation, and vice versa. Recently, we have derived analytical expressions to quantify these coupled particle displacements using no-slip boundary conditions for the matrix on the surfaces of the particles \cite{puljiz2016forces,puljiz2017forces}. %up to the order $\left(a/r_{ij}^{(0)}\right)^{\!4}$. In this expression, ${r}_{ij}^{(0)}=\left|\mathbf{r}_{ij}^{(0)}\right|$, and the distance vector $\mathbf{r}_{ij}^{(0)}=\mathbf{r}_i^{(0)}-\mathbf{r}_j^{(0)}$ refers to the initial state at $\mathbf{H}^\text{ext}=\mathbf{0}$. 
The resulting displacements for our two-particle system in response to the magnetic forces $\mathbf{F}_i$ are given by
\begin{equation}\label{displaceability_matrix}
	\mathbf{U}_i ={} \mathbf{\underline{M}}_{ii}\cdot\mathbf{F}_i + \mathbf{\underline{M}}_{i\neq j}\cdot\mathbf{F}_j,
\end{equation}
where $(i,j)\in\{(1,2),(2,1)\}$, while $\mathbf{\underline{M}}_{ii}$ and $\mathbf{\underline{M}}_{i\neq j}$ are called displaceability matrices \cite{puljiz2016forces,puljiz2017forces} (we mark tensors of second rank and mathematical matrices by an underscore).
Here, the corresponding expressions read
\begin{eqnarray}
	\mathbf{\underline{M}}_{ii} &={}& M_0\Bigg[ \mathbf{\underline{\hat{I}}} - \frac{15}{4} \left(\frac{a}{r_{ij}^{(0)}}\right)^{\!\!4} \mathbf{\hat{r}}_{ij}^{(0)}\mathbf{\hat{r}}_{ij}^{(0)} \Bigg], \label{M_ii}\\
	\mathbf{\underline{M}}_{i\neq j} &={}& M_0 \frac{3}{4} \frac{a}{r_{ij}^{(0)}}\Bigg[ (\mathbf{\underline{\hat{I}}}+\mathbf{\hat{r}}_{ij}^{(0)}\mathbf{\hat{r}}_{ij}^{(0)})  \notag\\
	&{}&\qquad\qquad\quad + 2\left(\frac{a}{r_{ij}^{(0)}}\right)^{\!\!2}\left(\frac{1}{3}\mathbf{\underline{\hat{I}}} -\mathbf{\hat{r}}_{ij}^{(0)}\mathbf{\hat{r}}_{ij}^{(0)} \right) \Bigg],\qquad\label{M_ij}
\end{eqnarray}
with $M_0=1/6\pi\mu a$, $\mathbf{\underline{\hat{I}}}$ the unity matrix, $\mathbf{\hat{r}}_{ij}^{(0)}=\mathbf{r}_{ij}^{(0)}/{r}_{ij}^{(0)}$, ${r}_{ij}^{(0)}=\left|\mathbf{r}_{ij}^{(0)}\right|$, and the distance vector $\mathbf{r}_{ij}^{(0)}=\mathbf{r}_i^{(0)}-\mathbf{r}_j^{(0)}$ refers to the initial state at $\mathbf{H}^\text{ext}=\mathbf{0}$. 
These displaceability matrices essentially contain all matrix-mediated elastic interactions between the two spheres mediated by the embedding elastic environment up to (including) order $\left(a/r_{ij}^{(0)}\right)^{\!4}$.
%The closer the particles come together, the higher the impact of higher-order terms in the inverse interparticle distances $r_{ij}^{(0)}$.

Naturally, the stronger the magnetic forces $\mathbf{F}_i$ on the particles, the larger the particle displacements, see Eq.~(\ref{displaceability_matrix}). Simultaneously, the magnetic forces $\mathbf{F}_i$ significantly increase with decreasing particle separation. Therefore, another iteration loop is necessary to calculate the final configuration at each magnitude of the external magnetic field \cite{puljiz2016forces,puljiz2017forces,menzel2017force}.

%In order to calculate the actual magnetic moments, first the susceptability $\mu_r-1$ of Ni in the required magnetic field interval [$0.0$, $60.8$] mT had to be determined. For this purpose, a magnetization curve from \dots was used, see Fig.~\ref{fig2}. Since the maximum magnitude of the externally applied magnetic field, $60.8$\,mT, was still in the approximately linear regime of the magnetization curve, we fitted Eq.~(\ref{magnetization_1}) via $\mu_r-1$, see Fig.~\ref{fig2}, and obtained $\mu_r=2.54$.

Solving Eqs.~(\ref{H_dipole})--(\ref{M_ij}) numerically by iteration as described, we can for a given external magnetic field calculate the resulting interparticle distance. Using a chi-square fit \cite{press1982numerical} to the experimental data points, we extracted in this way a local shear modulus $\mu=(226.0\pm2.8)\:\SI{}{\pascal}$ of the elastic matrix and an initial interparticle distance $r_{12}^{(0)}=(303.0\pm{0.1})\:\SI{}{\micro\metre}$ within the experimental error.
The resulting theoretical fitting curve for the interparticle distance $r_{12}/a$ as a function of the magnitude $B^\text{ext}$ of the external magnetic field is shown in Fig.~\ref{fig3}.

At an external magnetic field strength of $B^\text{ext}\approx\SI{61.7}{\milli\tesla}$, the linear elastic theory predicts a steep drop. At this point, the magnetic forces outmatch the linearly elastic restoring forces. Since the magnetic forces strongly increase with decreasing interparticle distance, they can grow significantly more strongly than the restoring elastic forces. Therefore, the separated state collapses. The particles are driven towards each other, until their steric volume exclusion hinders penetration when their surfaces virtually get into contact and basically touch each other. Experimentally, the field strength at which the collapse occurs could be located within the interval $B^\text{ext}\in]\SI{60.8}{\milli\tesla},\SI{65.1}{\milli\tesla}]$, in line with the prediction of our theory. 

Finally, we address the question of hysteresis within our linearly elastic description. In our situation this would imply that, starting from the collapsed state and subsequently reducing the magnitude of the field, the particles separate at a lower magnetic field strength than the one at which they collapsed. 
Such a behavior is conceivable in our picture. The magnitude of the restoring elastic forces in the collapsed state is independent of the strength of the external magnetic field. %It needs to be outmatched by the magnetic forces to drive the particles into this collapsed state or to maintain the collapsed state. 
However, the magnitude of the mutual magnetic interaction forces strongly depends on the actual distance between the two particles. 
It scales approximately with the quartic inverse particle separation. Therefore, a stronger external magnetic field is necessary to overcome the elastic barrier when the particles are well separated than to maintain the collapsed state when the particles have already approached. 

To calculate the value of $B^{\text{ext}}$ at which the separation occurs, we thus first determined the magnitude of the restoring elastic force in the collapsed state. This was performed by stepwise increasing an external force that drives the particles together, the magnitude of which not depending on the particle separation. The force at which we reach $r_{12}\approx2a$ identifies the strength of the elastic restoring forces. Then, artificially keeping the particles in the collapsed situation, we decreased $B^{\text{ext}}$ from a very high value until the attractive magnetic forces just balanced the previously determined elastic restoring forces. From this procedure, we obtained $B^\text{ext}\approx\SI{38.3}{\milli\tesla}$ for the separation of the particles. This detachment is indicated in Fig.~\ref{fig3} and, in this simplified picture, signals pronounced hysteresis.

The interparticle distance at which the collapse starts is not too far away from the last experimental data point, which is still well represented by the theory. Moreover, its magnitude of $r_{12}/a\approx3$ is still significantly larger than in the collapsed state ($r_{12}/a\approx2$). Therefore, we expect that our theory, despite its simplifications, particularly the dipolar picture and the restriction to linear elasticity, still captures the point of collapse reasonably well. The situation is different in the collapsed state, when the surfaces of the particles virtually touch each other. In this state, the material between the particles is strongly distorted, and nonlinear elastic effects certainly play a central role. Moreover, the magnetization across the interior of the particles becomes inhomogeneous, which challenges our reduced description in terms of magnetic dipoles. Therefore, in the collapsed state a nonlinear approach and a spatially resolved treatment of the magnetization are necessary to describe the behavior quantitatively correctly, see Sec.~\ref{xfem}. Before we address this issue, however, we demonstrate that a mapping onto significantly reduced models is possible in the present set-up.

\section{Mapping onto reduced dipole-spring models}
\label{dipole-spring}

\begin{figure}[t!]
{\includegraphics[width=5.2cm]{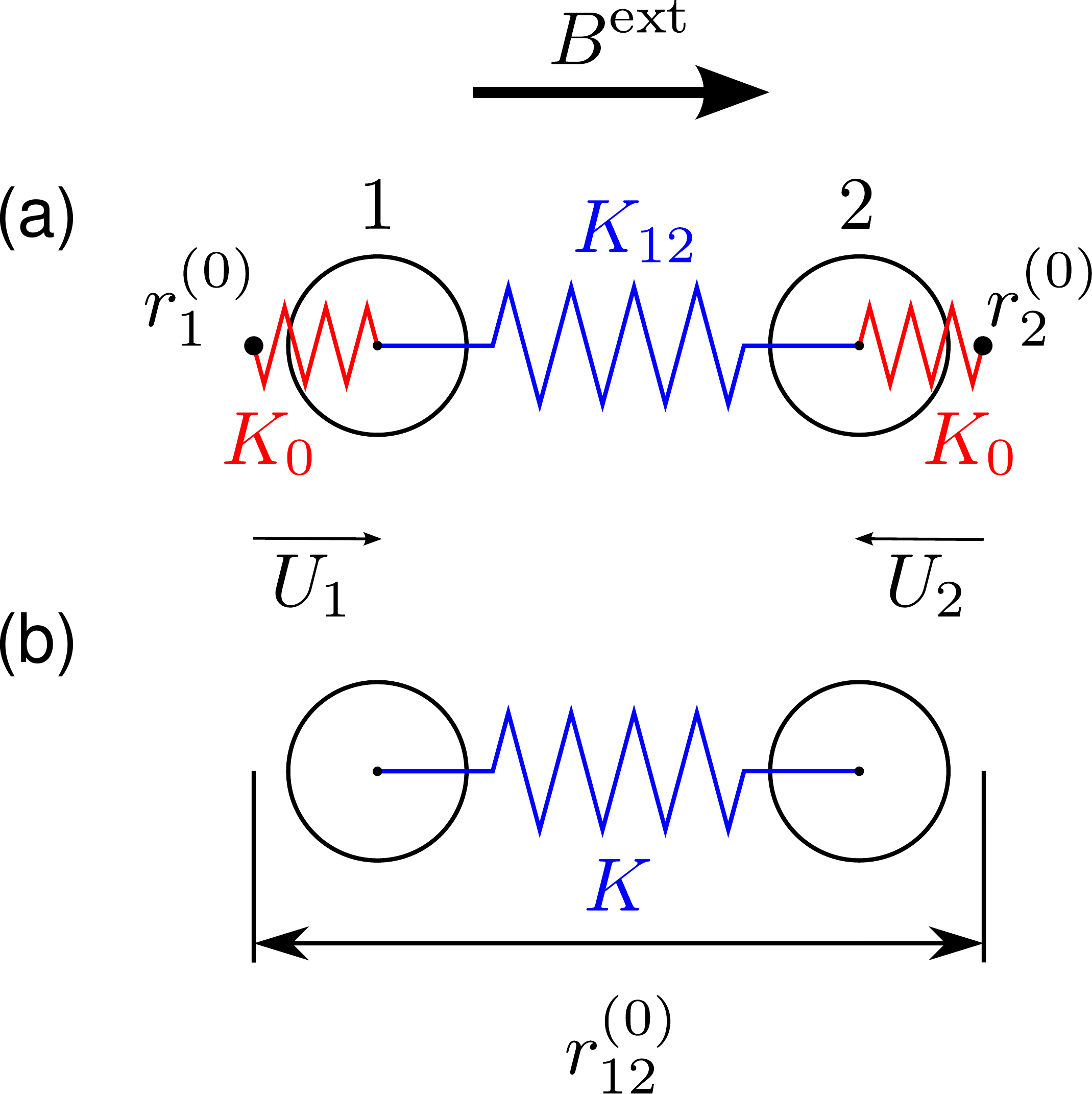}}
\caption{
Two reduced spring models that exploit the symmetry of the set-up (all vector quantities characterizing and affecting the particles are projected onto the symmetry axis along $\mathbf{B}^{\text{ext}}$). (a) Three-spring model. Each particle is anchored by one harmonic spring of spring constant $K_0$ to its ground-state position (red). Moreover, it is connected to the other particle by one additional harmonic spring of spring constant $K_{12}$ (blue). (b) One-spring model. The three-spring approach can further be reduced to a one-spring model involving only one effective harmonic spring of spring constant $K$ connecting the two particles.
}
\label{fig4}
\end{figure}

To the given order in the inverse particle separation and within the framework of linear elasticity theory, our approach in Sec.~\ref{linear-elastic} is exact concerning the treatment of the elastic polymer matrix. In previous approaches, simplified spring-like interactions had been introduced to model the matrix elasticity \cite{stepanov2008motion,annunziata2013hardening,pessot2014structural,tarama2014tunable,ivaneyko2015dynamic, allahyarov2015simulation,allahyarov2016dipole,pessot2016dynamic, cremer2017density,pessot2018tunable}. We now argue that in the present highly symmetric and simplified set-up the reduction to effective harmonic spring-like interactions is exact within the framework of linear elasticity theory. Moreover, the spring constants can be calculated as a function of the given parameters. 

%For comparison, we now want to consider a simplified situation, where the infinite elastic matrix around the spherical particles is replaced by one or more harmonic springs.

We consider the interparticle unit vector $\mathbf{\hat{r}}_{21}=-\mathbf{\hat{r}}_{12}$ to point along $\mathbf{B}^\text{ext}$. Then, by the symmetry of the set-up, all particle displacements, magnetic moments, and magnetic forces are oriented parallel to $\mathbf{\hat{r}}_{21}\|\mathbf{B}^\text{ext}$. Thus, we project all vector quantities by scalar multiplication onto this axis of symmetry. This leads to a scalar formulation of the theory in Sec.~\ref{linear-elastic} in terms of $U_i=\mathbf{U}_i\cdot\mathbf{\hat{r}}_{21}$, $m_i=\mathbf{m}_i\cdot\mathbf{\hat{r}}_{21}$, and $F_i=\mathbf{F}_i\cdot\mathbf{\hat{r}}_{21}$, $i\in\{1,2\}$. 
In this way, Eq.~(\ref{displaceability_matrix}) reduces to 
%\am{bitte explicit ausschreiben fuer Vektoren $(U_1,U_2)=...)$ als Matrixgleichung (Spaltenvektoren)}. 
%
\begin{eqnarray}
\lefteqn{\left(
\begin{array}{c}
U_1 \\[.5cm] U_2 
\end{array}
\right)
=}
\nonumber\\[.1cm]
&&
M_0
\left(
\begin{array}{cc}
1-\frac{15}{4}\left(\frac{a}{r_{ij}^{(0)}}\right)^{\!4} \;&\; \frac{3}{2} \frac{a}{r_{ij}^{(0)}}-\left(\frac{a}{r_{ij}^{(0)}}\right)^{\!3} \\[.3cm] 
\frac{3}{2} \frac{a}{r_{ij}^{(0)}}-\left(\frac{a}{r_{ij}^{(0)}}\right)^{\!3} \;&\; 1-\frac{15}{4}\left(\frac{a}{r_{ij}^{(0)}}\right)^{\!4}
\end{array}
\right)
\cdot
\left(
\begin{array}{c}
F_1 \\[.5cm] F_2 
\end{array}
\right).
\nonumber\\
&&
\label{project1d}
\end{eqnarray}

\subsection{Three-spring model}

Illustratively, the matrix term in Eq.~(\ref{project1d}) suggests to introduce three harmonic springs to model the elastic situation, see Fig.~\ref{fig4}~(a). 
On the one hand, its diagonal contains the zeroth-order contributions in the particle separation that arise already when one single particle is displaced against the surrounding elastic matrix. The corresponding counteracting force by the surrounding elastic matrix makes us introduce two harmonic springs of constant $K_0$ that anchor each particle to its ground state position.
The coefficient $K_0$ further contains a fourth-order correction that results from the displacement field induced in the matrix by one particle when this field is ``reflected'' by the rigidity of the other particle \cite{puljiz2016forces,puljiz2017forces}.
On the other hand, the off-diagonal entries describe the displacements of the particles due to the forces exerted on the respective other particle. These particle interactions are mediated by the elastic matrix. Consequently, we introduce another harmonic spring of constant $K_{12}$ coupling the two particles to each other. 
Together, this spring-model leads to the elastic potential energy
\begin{equation}\label{V_el_3}
	W^\text{el}={} \frac{1}{2}K_0\left(U_1^2+U_2^2\right)+\frac{1}{2}K_{12}\left(U_1-U_2\right)^2.
\end{equation}
%
%\am{alles $V$ oder alles $W$} 
From this expression, the elastic restoring forces $F_i^{\text{el}}$, $i\in\{1,2\}$, follow as 
\begin{equation}\label{F_el3}
	F_i^\text{el}={}-\frac{\partial}{\partial U_i}W^\text{el}. 
	%=\mpu{-K_0 U_i - K_{12}(U_i - U_j) }.
\end{equation}

\begin{widetext}
To calculate the spring constants $K_0$ and $K_{12}$ from our linear elasticity theory, we invert Eq.~(\ref{project1d}) and obtain, up to (including) order $\left(a/r_{ij}^{(0)}\right)^{4}$, %\am{bitte einfuegen}
\begin{equation}
\left(
\begin{array}{c}
F_1 \\[.5cm] F_2 
\end{array}
\right)
=
\frac{1}{M_0}
\left(
\begin{array}{cc}
1+\frac{9}{4}\left(\frac{a}{r_{ij}^{(0)}}\right)^{\!2}+\frac{93}{16}\left(\frac{a}{r_{ij}^{(0)}}\right)^{\!4} 
\;&\; -\frac{3}{2} \frac{a}{r_{ij}^{(0)}}-\frac{19}{8}\left(\frac{a}{r_{ij}^{(0)}}\right)^{\!3} \\[.4cm] 
-\frac{3}{2} \frac{a}{r_{ij}^{(0)}}-\frac{19}{8}\left(\frac{a}{r_{ij}^{(0)}}\right)^{\!3} 
\;&\; 1+\frac{9}{4}\left(\frac{a}{r_{ij}^{(0)}}\right)^{\!2}+\frac{93}{16}\left(\frac{a}{r_{ij}^{(0)}}\right)^{\!4} 
\end{array}
\right)
\cdot
\left(
\begin{array}{c}
U_1 \\[.5cm] U_2 
\end{array}
\right).
\label{project1dforces}
\end{equation}
This equation lists the forces $F_1$ and $F_2$ that need to be imposed onto the particles to achieve their displacements $U_1$ and $U_2$. Corresponding expressions for the forces as functions of the displacements have been derived directly, i.e., without the intermediate inversion, already in Ref.~\citenum{phan1994load} and agree with ours to the given order. 
\end{widetext}

In a stationary state, the forces $F_1$ and $F_2$ in Eq.~(\ref{project1dforces}) imposed on the particles need to be equal in magnitude but oppositely oriented to the forces $F_1^{\text{el}}$ and $F_2^{\text{el}}$ exerted onto the particles by the elastic matrix, respectively. Comparing Eqs.~(\ref{F_el3}) and (\ref{project1dforces}), we read off the spring constants
\begin{eqnarray}
K_0 &=& \frac{1}{M_0}\bigg[1-\frac{3}{2} \frac{a}{r_{ij}^{(0)}}+\frac{9}{4}\left(\frac{a}{r_{ij}^{(0)}}\right)^{\!2}-\frac{19}{8}\left(\frac{a}{r_{ij}^{(0)}}\right)^{\!3}
\nonumber\\
&&{}+\frac{93}{16}\left(\frac{a}{r_{ij}^{(0)}}\right)^{\!4}\bigg],  \\
K_{12} &=& \frac{1}{M_0}\bigg[\frac{3}{2} \frac{a}{r_{ij}^{(0)}}+\frac{19}{8}\left(\frac{a}{r_{ij}^{(0)}}\right)^{\!3}\bigg].
\end{eqnarray}

Overall, in this dipole-spring model, the total energy characterizing a certain configuration is given by
\begin{equation}\label{V_tot}
	W^\text{tot}={}W^\text{mag}+W^\text{el}.
\end{equation}
Here, the magnetic contribution follows from Eq.~(\ref{Wmag}) and is calculated in the same way as described in Sec.~\ref{linear-elastic}. 

To find the state of our system for a given $B^\text{ext}$, we minimize $W^\text{tot}$ using a simple relaxational method \cite{press1982numerical}. In this way, we can stepwise increase and decrease $B^\text{ext}$, inserting as an initial condition after each modification of $B^\text{ext}$ the previously relaxed state of the system. As a result, we naturally obtain the hysteresis loop in the particle separation, depicted in Fig.~\ref{fig3} by the dashed line. 
There, the deviations from the solid line obtained from the calculation using linear elasticity theory in Sec.~\ref{linear-elastic} can be traced back to our matrix inversion from Eq.~(\ref{project1d}) to Eq.~(\ref{project1dforces}). As the overall formalism, this inversion is exact only to (including) order $\left(a/r_{ij}^{(0)}\right)^{4}$. If, instead, we invert Eqs.~(\ref{F_el3}), $i\in\{1,2\}$, and obtain the spring constants by comparison with Eq.~(\ref{project1d}), the two curves collapse. Thus the deviations in Fig.~\ref{fig3} between the direct calculation using continuum elasticity theory as in Sec.~\ref{linear-elastic} and our dipole-spring reduction represent higher-order effects. %\am{stimmt das so?}

\subsection{One-spring model} 

Further exploiting the symmetry of our geometry, it is straightforward to additionally reduce Eq.~(\ref{V_el_3}) to the situation of one effective spring of constant $K$ connecting the two particles, see Fig.~\ref{fig4}~(b). Mirror symmetry of the set-up dictates the relations $U_1={}-U_2$, $F_1={}-F_2$, as well as $m_1=m_2=m$ for the magnetic moments.
Defining $U=U_1={}-U_2$, we obtain
\begin{equation}\label{V_el}
%	W^\text{el}={} \frac{1}{4}K\left(U_1-U_2\right)^2={}K U_1^2={}K U_2^2,
W^\text{el}={} \frac{1}{2}KU^2,
\end{equation}
with the effective spring constant $K=2K_0+4K_{12}$ and the elastic forces on the two particles given by $\mp\partial W^\text{el}/2\partial U$.
%\am{ich wuerde dann hier das $K$ nicht mehr explizit hinschreiben, es ergibt sich ja aus den anderen ausdruecken}
%\mpu{Aber man sollte vielleicht darauf hinweisen, dass die Kraft pro Partikel jetzt nur noch $-1/2\partial_U W^\text{el}$ ist}
Naturally, we obtain the same hysteresis loop in Fig.~\ref{fig3} after rewriting the theory into the effective one-spring model in Eq.~(\ref{V_el}).
To the given order, it contains the same information as the linearly elastic approach in Sec.~\ref{linear-elastic}, therefore representing an effective reduction of the theory.

\begin{figure}
\centerline{\includegraphics[width=\columnwidth]{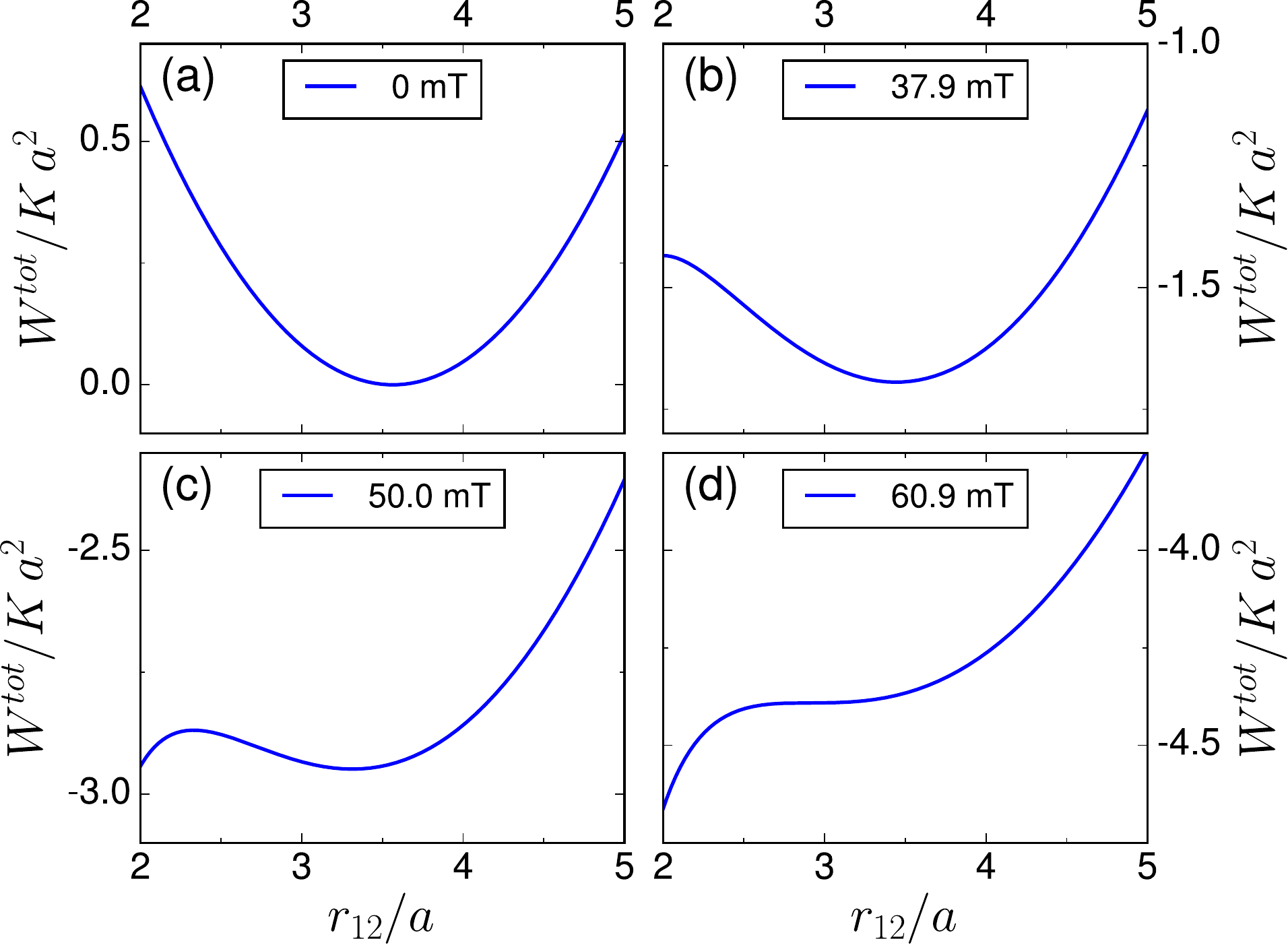}}
\caption{
Total energy $W^\text{tot}$ for the one-spring dipole-spring model as a function of the interparticle distance $r_{12}$ for increasing amplitude $B^\text{ext}$ of the external magnetic field.
(a) At $B^\text{ext}=\SI{0}{\tesla}$, $W^\text{tot}=W^\text{el}$ is given by Eq.~(\ref{V_el}), with one minimum at $r_{12}^{(0)}/a\approx{3.57}$.
(b) Starting from the threshold field $B^\text{ext}\approx{\SI{37.9}{\milli\tesla}}$, a second local minimum starts to develop at $r_{12}/a\approx2$. 
%Therefore, a collapsed state can be maintained for weaker magnetic field strengths then were necessary for creating the collapsed state, see Fig.~\ref{fig3}.
(c) With increasing $B^\text{ext}$, the minimum at $r_{12}/a\approx2$ deepens, whereas the minimum for $r_{12}/a\gtrsim3$ flattens out.
(d) Finally, for $B^\text{ext}\gtrsim{\SI{60.9}{\milli\tesla}}$, only the minimum at $r_{12}/a\approx2$ remains.
In the absence of thermal fluctuations and external perturbations, the system cannot cross the energetic barrier between the two local minima at intermediate field strengths ${\SI{37.9}{\milli\tesla}}\lesssim B^\text{ext} \lesssim{\SI{60.9}{\milli\tesla}}$. Only when one of the two minima vanishes at one of the two threshold strengths, a jump to the other minimium can occur, which explains the hysteresis loop in Fig.~\ref{fig3}. 
}
\label{fig5}
\end{figure}

Using the total energy of the system $W^\text{tot}$ in Eq.~(\ref{V_tot}) together with the elastic energy of the effective one-spring model in Eq.~(\ref{V_el}), we can readily illustrate the origin of the hysteretic behavior. Figure~\ref{fig5} shows the shape of the total energy for increasing amplitude of the external magnetic field. 

At $B^\text{ext}=\SI{0}{\milli\tesla}$, the overall energy features one minimum at a distance $r_{12}/a=r_{12}^{(0)}/a\approx{3.57}$ as displayed in Fig.~\ref{fig5}~(a). Starting from $B^\text{ext}\gtrsim{\SI{37.9}{\milli\tesla}}$, 
%\am{unteres ``Sprung''feld, am besten alles auf 1 Nachkommastelle fuer diese Felder, falls moeglich, auch im Text vorher und in Fig.5...} 
see Fig.~\ref{fig5}~(b), a second minimum begins to develop at the distance of close approach $r_{12}/a\approx 2$. As depicted in Fig.~\ref{fig5}~(c), two local minima in the overall energy then coexist over a broader range of external field amplitudes. Only around $B^\text{ext}\gtrsim{\SI{60.9}{\milli\tesla}}$, the initial minimum has vanished and the one remaining minimum at $r_{12}/a\approx2$ represents the collapsed state, see Fig.~\ref{fig5}~(d). 

If thermal fluctuations and external perturbations are negligible, the energetic barrier separating the two minima in the range of magnetic fields ${\SI{37.9}{\milli\tesla}}\lesssim B^\text{ext} \lesssim{\SI{60.9}{\milli\tesla}}$ cannot be crossed. Then, the state of the system in this regime depends on its history. If the external field is increased and the particles had previously been well separated, the system is trapped in the minimum at $r_{12}/a\gtrsim3$. Only at the threshold amplitude $B^{\text{ext}}\approx{\SI{60.9}{\milli\tesla}}$, where this minimum vanishes, see Fig.~\ref{fig5}~(d), the particles collapse to the state of $r_{12}/a\approx2$. Vice versa, upon then decreasing the field amplitude, the particles only re-separate at the lower threshold amplitude $B^{\text{ext}}\approx{\SI{37.8}{\milli\tesla}}$. There, the minimum at $r_{12}/a\approx2$ vanishes, see Fig.~\ref{fig5}~(b). In this way, the hysteresis loop in Fig.~\ref{fig3} can be directly read off from the energy curves in Fig.~\ref{fig5}. 
%\am{bitte auch arXiv in den refs. updaten}

\section{Results from finite-element simulations}
\label{xfem}
To test our theoretical conclusions and the consequences of our approximations, we performed additional FE simulations. They
%In this section, the numerical solution of the discussed two particle problem by FE simulations is presented. These simulations 
are based on a coupled magneto-mechanical continuum formulation including nonlinear contributions to the magnetic and elastic properties of the magnetic particles and elastic matrix, respectively \cite{kalina2016microscale,Kalina2017}. In addition to the theoretical approaches above, the magnetic %and mechanical fields 
field is spatially resolved also within the two particles.\\

\subsection{Continuum approach}

We consider a piece of the material described above %body represented by the configuration $\mathcal B \in \mathbb R^3$ 
with density distribution $\varrho(\mathbf{r})$, containing the two spherical particles. %and volume $V$. %at time $t$. 
%\am{no time dependence here, Volumen kommt auch nicht mehr vor...}
%The behavior of this body is described by a nonlinear boundary value problem (BVP).
%\am{can we skip the term bvp here? not needed afterwards...} 
In the stationary case, the magnetic part of the %BVP 
corresponding coupled continuum formulation is
given by the two Maxwell equations 
\begin{alignat}{5}	
	\nabla \cdot \mathbf B^\text{in}(\mathbf{r}) &= 0 &&\quad\text{with}\quad &\mathbf{\hat{n}}(\mathbf{r}) \cdot \llbracket \mathbf B^\text{in}(\mathbf{r}) \rrbracket &= 0 &&\;\text{on } \mathcal S^\text{d} &&\quad  \label{eq:Gauss}
\end{alignat}
and
\begin{alignat}{5}	
	\nabla \times \mathbf H^\text{in}(\mathbf{r}) &= \mathbf 0 &&\quad\text{with}\quad &\mathbf{\hat{n}}(\mathbf{r}) \times \llbracket \mathbf H^\text{in}(\mathbf{r}) \rrbracket &= \mathbf 0 &&\;\text{on }\mathcal S^\text{d}. % &&\quad\text{,} 
	\label{eq:Ampere}
\end{alignat}
In these expressions, \mbox{$\mathbf B^\text{in}(\mathbf{r})$} and \mbox{$\mathbf H^\text{in}(\mathbf{r})$} denote the local magnetic flux density and magnetic field inside the material, respectively, both in the magnetic particles and in the elastic matrix \cite{jackson1962classical}. They %local fields  $\mathbf B^\text{in}$ and $\mathbf H^\text{in}$ 
are connected via %by the linking equation 
\mbox{$\mathbf B^\text{in}(\mathbf{r}) = \mu_0 \left(\mathbf H^\text{in}(\mathbf{r}) + \mathbf M(\mathbf{r})\right)$}, with $\mathbf{M}(\mathbf{r})$ describing the magnetization field. $\mathcal S^\text{d}$ denotes a surface of discontinuity, here the interface between the particles and the elastic matrix. The brackets \mbox{$\llbracket\cdot\rrbracket$} quantify the jump of the contained quantity across $\mathcal S^\text{d}$, while \mbox{$\mathbf{\hat{n}}(\mathbf{r})$} corresponds to the unit normal vector on $\mathcal S^\text{d}$. %The local fields  $\mathbf B^\text{in}$ and $\mathbf H^\text{in}$ are connected by the linking equation \mbox{$\mathbf B^\text{in} = \mu_0 (\mathbf H^\text{in} + \mathbf M)$}. 
%\am{auch wenn es nervig aussieht, ich denke, hier sollten wir explizit die $(\mathbf{r})$-Abh\"angigkeiten mitschleppen im Gegensatz zu den vorherigen Abschnitten, damit die Feldbeschreibung deutlich wird}

The magnetic fields cause magnetic coupling terms that enter the mechanical part of the problem, %BVP, 
e.g., %\am{gibt es noch mehr au{\ss}er diesem $f^m$?} \textit{\Karl{es gibt noch die Momentendicte $\mathbf c^\text{mag}$ und die Leistungsdichte $P^\text{mag}$}} 
the magnetic body force density \mbox{$\mathbf f^\text{mag}(\mathbf{r})=\left(\nabla \mathbf B^\text{in}(\mathbf{r})\right)^\text{T}\cdot \mathbf M(\mathbf{r})$}. Here, the superscript $^\text{T}$ denotes the transpose. 
Since $\mathbf f^\text{mag}(\mathbf{r})$ can be expressed as the divergence of the magnetic stress tensor \mbox{$\mathbf{\underline{\boldsymbol{\sigma}}}^\text{mag}(\mathbf{r})$}, the balance of linear momentum is given by the relation \cite{Groot1972,Eringen1990,kalina2016microscale,Kalina2017}
\begin{align}
		\nabla \cdot \mathbf{\underline{\boldsymbol{\sigma}}}^\text{tot}(\mathbf{r}) = \boldsymbol{0} \quad\text{with}\quad &\mathbf{\hat{n}}(\mathbf{r}) \cdot \llbracket \mathbf{\underline{\boldsymbol{\sigma}}}^\text{tot}(\mathbf{r}) \rrbracket = \boldsymbol{0} \;\text{on } \mathcal S^\text{d}. \label{eq:LinMom}
\end{align}	
%
%Therein, 
Here, the total stress tensor $\mathbf{\underline{\boldsymbol{\sigma}}}^\text{tot}(\mathbf{r})$ is defined as the sum of the mechanical and the magnetic stress tensors, \mbox{$\mathbf{\underline{\boldsymbol{\sigma}}}^\text{tot}(\mathbf{r}) = \mathbf{\underline{\boldsymbol{\sigma}}}(\mathbf{r}) + \mathbf{\underline{\boldsymbol{\sigma}}}^\text{mag}(\mathbf{r})$}. A detailed discussion of the complete magneto-mechanical field equations is given, e.g., in Refs.~\citenum{Groot1972,Eringen1990,Kankanala2004,Ogden2011book}. 

The constitutive behavior of the considered particle-matrix system can be described by a specific free-energy density (per mass) $\Psi$, split into a magnetic and a mechanical part \cite{PonteCastaneda2011,Javili2013,metsch2016numerical, kalina2016microscale,Kalina2017,romeis2017theoretical,Danas2017}
\begin{equation}
\Psi\!\left(\mathbf{\underline{F}}(\mathbf{r}), \mathbf H^\text{in}(\mathbf{r})\right) = \Psi^\text{mag}\!\left(\mathbf H^\text{in}(\mathbf{r})\right) + \Psi^\text{el}\big(\mathbf{\underline{\mathbf{F}}}(\mathbf{r})\big). 
\end{equation}
In this expression, $\mathbf{\underline{F}}(\mathbf{r})$ denotes the deformation gradient tensor. Hence, the constitutive relations
\begin{align}\label{M_psimag}
	\mathbf M(\mathbf{r}) &= - \frac{\varrho(\mathbf{r})}{\mu_0} \frac{\partial\, \Psi^\text{mag}\!\left(\mathbf H^\text{in}(\mathbf{r})\right)}{\partial \mathbf H^\text{in}(\mathbf{r})} 
\end{align}
and
\begin{align}
		 \mathbf{\underline{\boldsymbol{\sigma}}}(\mathbf{r}) &= \varrho(\mathbf{r}) \mathbf{\underline{F}}(\mathbf{r}) \cdot \frac{\partial\, \Psi^\text{el}\big(\mathbf{\underline{F}}(\mathbf{r})\big)}{\partial \mathbf{\underline{F}}^\text{T}(\mathbf{r})} - \frac{\mu_0}{2} \left(\mathbf M(\mathbf{r}) \cdot \mathbf M(\mathbf{r})\right)\mathbf{\underline{\hat{I}}} 
\end{align}
%
%\am{delta's f\"ur Funktionalableitungen okay?} \textit{\Karl{Ich würde sagen $\Psi$ ist kein Funktional. Ein Funktional ist meiner Meinung nach erst das Potential 
%\begin{align*}
%	\Pi^\text{int}\left\{\mathbf{\underline{\boldsymbol F}}(\mathbf{r}), \mathbf H^\text{in}(\mathbf{r})\right\} = \int_{\mathcal B} \varrho\Psi(\mathbf{\underline{\boldsymbol F}}, \mathbf H^\text{in}) - \frac{1}{2} \mu_0 \mathbf H^\text{in} \cdot \mathbf H^\text{in} dV.
%\end{align*}
%Durch die Integration über das Volumen wird hier für jede Feldverteilung $\boldsymbol F(\mathbf r)$ und $\mathbf H^\text{in}(\mathbf r)$ ein inneres Potential für das betrachtete Gebiet bestimmt. Nach meiner Auffassung spielt erst dann der Funktions-Charakter, d.h, die Abhängigkeit von $\mathbf r$, eine Rolle. Sollte man bei $\Psi$ die Abhängigkeit dann lieber ohne geschweifte Klammern angeben?}}
can be found from the evaluation of the second law of thermodynamics, see, e.g., Refs.~\citenum{Kankanala2004,Ogden2011book}. 
% If the mechanical behavior of both particles and matrix is described by an elastic neo-Hooke model and the specific free energy function $\Psi = \Psi(\boldsymbol F, \mathbf H^\text{in})$ is given in terms of the deformation gradient $\boldsymbol F$ and the local magnetic field $\mathbf H^\text{in}$, $\boldsymbol \sigma^\text{tot}$ can be expressed as

Eq.~(\ref{eq:Langevin}) already defines a relation between the magnetization and the magnetic field within the particles, while $\mathbf{M}(\mathbf{r})=\mathbf{0}$ within the elastic matrix, so that we choose $\Psi^\text{mag}$ in Eq.~(\ref{M_psimag}) accordingly. % is chosen such that the magnetization $\mathbf M(\mathbf{r})$ is   described by the Langevin function and equal to zero in the matrix.
We specify the mechanical part $\Psi^\text{el}$ in both components, the particles and the elastic matrix, following an elastic Neo-Hookean model \cite{kalina2016microscale,Kalina2017}. Accordingly, $\mathbf{\underline{\boldsymbol{\sigma}}}^\text{tot}(\mathbf{r})$ is expressed as
\begin{eqnarray}
		\mathbf{\underline{\boldsymbol{\sigma}}}^\text{tot}(\mathbf{r}) 
		&= &\frac{1}{J(\mathbf{r})}\bigg[ \mu(\mathbf{r})\left(\mathbf{\underline{b}}(\mathbf{r}) - \mathbf{\underline{\hat{I}}}\right) 
		\nonumber\\[.05cm]
		&& \qquad\qquad {}+ \frac{\mu(\mathbf{r})\nu(\mathbf{r})}{1-2\nu(\mathbf{r})}\left(\left(J(\mathbf{r})\right)^2-1\right)\,\mathbf{\underline{\hat{I}}} \bigg] \nonumber\\[.15cm]
		&&{}+ \mathbf B^\text{in}(\mathbf{r})\,  \mathbf H^\text{in}(\mathbf{r}) - \frac{\mu_0}{2} \left(\mathbf H^\text{in}(\mathbf{r}) \cdot \mathbf H^\text{in}(\mathbf{r})\right) \,\mathbf{\underline{\hat{I}}},
		\nonumber\\
\end{eqnarray}	
where $\mathbf{\underline{b}}(\mathbf{r})$ is the left Cauchy--Green deformation tensor, $J(\mathbf{r}) = \det \mathbf{\underline{F}}(\mathbf{r})$ denotes the Jacobi determinant of the deformation gradient tensor, $\mu(\mathbf{r})$ is the local elastic shear modulus, and $\nu(\mathbf{r})$ is the local Poisson ratio. Within the elastic matrix, the material is characterized by the shear modulus $\mu$ determined in Sec.~\ref{linear-elastic} and the Poisson ratio $\nu=0.49$, corresponding to a nearly incompressible material. %According to Ref. \citenum{Dubbel2011}, we choose the parameters $\mu=\SI{75.19}{\giga\pascal}$ and $\nu=0.31$ within the magnetizable nickel particles. 
Inside the magnetizable nickel particles, we set the parameters of elasticity to $\mu=\SI{80.77}{\giga\pascal}$ and $\nu=0.3$ (which is of the order of magnitude of the values listed, e.g., in Ref.~\citenum{Dubbel2011}). The exact numerical values for the nickel particles are not significant because of their much higher mechanical stiffness relatively to the polymeric matrix.  
%\textit{\Karl{Ich habe noch mal nach Werten für $E$ und $\nu$ bzw. $\mu$ und $\nu$ in der Literatur geschaut und dementsprechend minimal angepasst. Das hat auf die Rechnungen natürlich keinerlei Auswirkungen, da die Steifigkeit der Partikel im Vergleich zur Matrix so hoch ist, dass quasi keine Deformation auftritt. Ich hoffe das ist ok so?}}

\subsection{Numerical solution}

In our FE simulations, we address a cuboid region %\mbox{$\mathcal R$} 
%\am{ich denke, die Definition \mbox{$\mathcal R$} braucht man auch nicht unbedingt?}
of the system with fixed boundary %\mbox{$\partial \mathcal R$} 
and dimensions $22a \times 10a \times 10a$. Exploiting the symmetries of the set-up, only one eighth of this region needs to be evaluated explicitly with appropriate boundary conditions.    
Major challenges result from the underlying bifurcation scenario and the extremely high degrees of deformation of the elastic matrix between the two particles in the virtual touching state. As a consequence, it was not possible to solve the problem in a direct way using the fully coupled FE solution scheme \cite{kalina2016microscale,Kalina2017}. Therefore, we separately calculated the magnitudes $F^\text{mag}$ of the attractive magnetic force and the magnitude $F^\text{el}$ of the counteracting mechanical restoring force on each particle for decreasing particle distances. Afterwards, we identified the states in which these two forces balance each other. %\am{okay?}

\begin{figure}
\centerline{\includegraphics{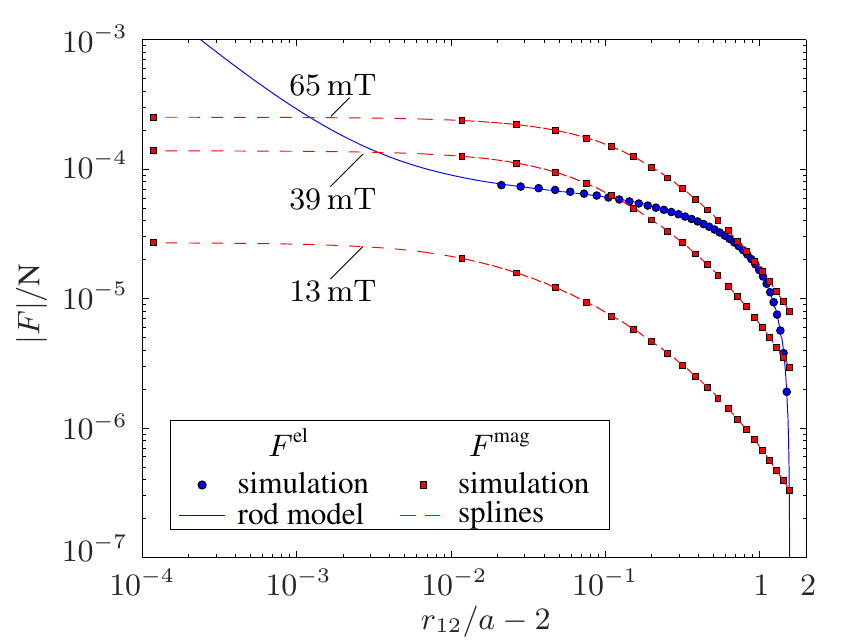}}
\caption{FE simulation results for the magnitudes $F^\text{mag}$ and $F^\text{el}$ of the attractive magnetic and counteracting elastic restoring forces, respectively, at various separation distances $r_{12}$ between the two particles. Values for $F^\text{mag}$ are interpolated using cubic splines, and example curves for magnetic field amplitudes \mbox{$B^\text{ext} = \SI{ 13}{\milli\tesla}$, $\SI{39}{\milli\tesla}$, and $\SI{65}{\milli\tesla}$} are shown. Values for the mechanical restoring force $F^\text{el}$ due to the deformation of the elastic matrix are fitted by an elastic rod model according to Eq.~(\ref{eq:RodModel}). Actual distances between the two particles for a given value of \mbox{$B^\text{ext}$} are identified from the intersections between the curves for the two different forces.}
\label{fig:Force}
\end{figure}

First, for each evaluated magnitude of the external magnetic flux density $\mathbf B^\text{ext}$, we determined $F^\text{mag}$ for $22$ different particle distances \mbox{$2.001 \le r_{12}/a \le 3.565$}. For this purpose, in total $22$ different FE meshes were generated. $F^\text{mag}$ is calculated for each distance by fixing the center positions of the particles. We then determine the necessary mechanical force to keep the particle positions fixed within a coupled magneto-mechanical FE simulation %\am{okay so? ich denke, der Ausdruck bearing force sagt den Physikern nicht allzu viel, deshalb habe ich ihn rausgeworfen.}
\cite{metsch2016numerical,kalina2016microscale,Kalina2017}. The magnitude $B^\text{ext}$ of the external magnetic field is increased in $100$ increments from $\SI{0}{\milli\tesla}$ to $\SI{65}{\milli\tesla}$. %DEN SATZ VERSTEHE ICH NICHT; ER ERSCHEINT MIR IM MOMENT NICHT UEBERWICHTIG, DA ICH DEN REST SCHON VERSTEHE; HABE IHN DAHER AUSKOMMENTIERT. As a result of the large ratio between the stiffness of particles and matrix, the distortion of $F^\text{mag}$ due to the mechanical influence of the matrix is negligible \cite{Lux2016}. 
The obtained values for $F^\text{mag}$ for each considered magnitude $B^\text{ext}$ are then interpolated between the evaluated distances $r_{12}$ using cubic splines. Example curves for $B^\text{ext} = \SI{ 13}{\milli\tesla}$, $\SI{39}{\milli\tesla}$, and $\SI{65}{\milli\tesla}$ are shown in Fig.~\ref{fig:Force}.  

Second, to determine $F^\text{el}$, the centers of the non-magnetized particles are moved towards each other within another FE simulation that addresses this purely mechanical problem of nonlinear elasticity. Since it is not possible to simulate interparticle distances arbitrarily close to $r_{12}/a \approx 2$ because of the strong distortion of the FE meshes, a reduced fitting function of sufficient physical significance needs to be employed to extrapolate $F^\text{el}$ in this regime. We chose the heuristic rod model introduced in Ref.~\citenum{biller2014modeling} for this purpose.
%\textit{\Karl{Ich hab noch mal nachgeschaut, in der Veröffentlichung wird es als 'heuristic scheme' eingeführt. Ich würde es dann aber hier später trotzdem als rod model bezeichnen.}}
It replaces the action of the elastic matrix by five nonlinearly elastic rods of lengths $l_{0,i}$ and radii $r_{0,i}$ in the undeformed configuration ($i=1,...,5$). This rod model allows to calculate $F^\text{el}$ analytically and to extrapolate the simulation results down to $r_{12}/a \to 2$.
As a benefit, the constitutive elastic behavior of the rods can be described by the same Neo-Hookean material model that we use for the elastic matrix in the FE simulation. In this framework, $F^\text{el}$ is given by
\begin{align}
	&F^\text{el} = \sum_{i=1}^5 \pi r_{0,i}^2 \left[ \mu\left(\lambda_{\text{l},i} - \frac{1}{\lambda_{\text{l},i}}\right) + \frac{\beta}{2}\left(\lambda_{\text{l},i} \lambda_{\text{t},i}^4 - \frac{1}{\lambda_{\text{l},i}}\right)\right] \nonumber\\
	&\text{with} \quad \lambda_{\text{t},i} = -\frac{\mu}{\beta \lambda_{\text{l},i}^2}+\sqrt{\frac{\mu^2}{\beta^2\lambda_{\text{l},i}^4}+\frac{2\mu}{\beta \lambda_{\text{l},i}^2} + \frac{1}{\lambda_{\text{l},i}^2}}\;. \label{eq:RodModel}
\end{align}
Here, $\lambda_{\text{l},i} = l_i/l_{0,i}$ and $\lambda_{\text{t},i}=r_i/r_{0,i}$ denote the deformation ratios of the rods in the directions parallel (longitudinal) and perpendicular (transversal) to the rod axes, respectively, with $l_i$ and $r_i$ the corresponding rod dimensions in the deformed state. %The Lam{\'e} parameter $\beta$ is expressible by 
Moreover, \mbox{$\beta= {2\mu\nu}/({1-2\nu})$}, where $\mu$ is the elastic shear modulus and $\nu$ is the Poisson ratio of the elastic polymer matrix. 
%HABE DEN EINDRUCK, DER SATZ IST NICHT NOETIG: The lengths $l_{0i}$ of the rods result from the initial position of the particles and the length $L_0 = 22a$ of the region $\mathcal R$. To describe the complex behavior of the surrounding matrix material, 
This effective rod model is fitted to the FE-simulation results using the radii $r_{0,i}$ as fit parameters, with the corresponding values listed in Tab.~\ref{tab1}. 
\begin{table}[b!]
\small
  \caption{Parameters of the rod model Eq.~(\ref{eq:RodModel}): lengths $l_{0,i}$ and fitted radii $r_{0,i}$ of the five rods.}
  \label{tab1}
  \begin{tabular*}{0.49\textwidth}{@{\extracolsep{\fill}}cccccc}
    \hline
    $i$ & $1$ & $2$ & $3$ & $4$ & $5$ \\
    \hline
    $l_{0,i}$ & $r_{12}\!-\!2a$ & $r_{12}\!-\!a$ & $r_{12}\!-\!a$ & $(20a\!-\!r_{12})/2$ & $(20a\!-\!r_{12})/2$ \\
    $r_{0,i}/\SI{}{\micro\metre}$ & $2.80$ & $72.17$ & $72.17$ & $273.15$ & $273.15$ \\
    \hline
  \end{tabular*}
\end{table}
Fig.~\ref{fig:Force} displays the results for $F^\text{el}$ from the FE simulation together with the fitted rod model according to Eq.~(\ref{eq:RodModel}). 
With a maximum error of $\SI{3.3}{\percent}$, the model shows acceptable deviations from the FE simulation. %KOMMT UNTEN: As depicted in Fig.~\ref{fig3}, the model is sufficient to reproduce the experimental results accurately. 
Furthermore, it describes the expected behavior of %polymers due to high compression in the extrapolated region , i.\,e., $|\sigma| \to \infty$ 
diverging restoring force for $r_{12}\to 2a$.
\begin{figure}
\centerline{\includegraphics{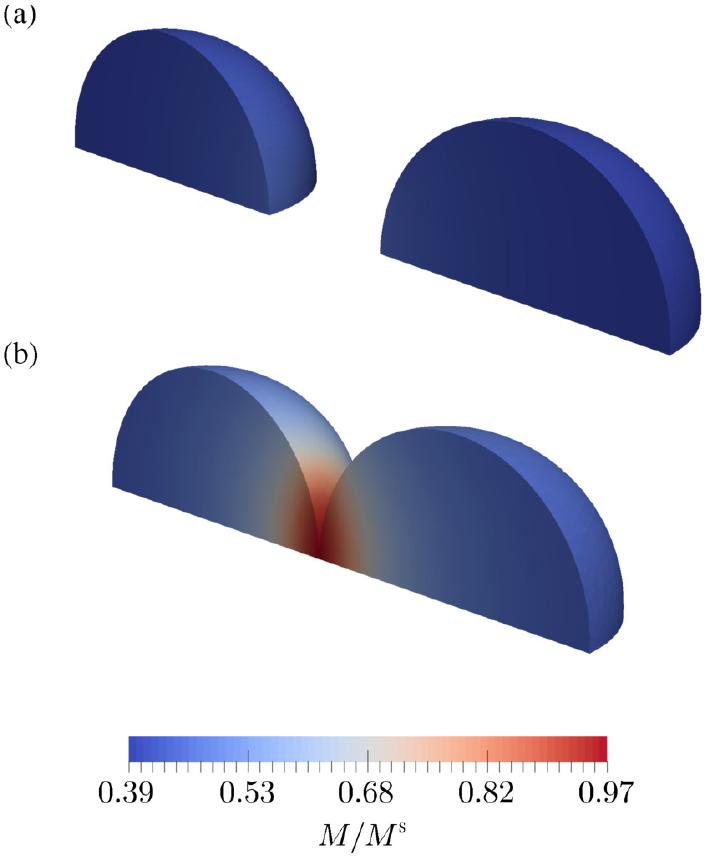}}
\caption{Color plot of the magnitudes of the spatially resolved magnetization field within the spherical particles (a) for $B^\text{ext}=\SI{63.05}{\milli\tesla}$ shortly before the collapse and (b) for \mbox{$B^\text{ext}=\SI{65.00}{\milli\tesla}$} after the collapse in the state of virtual touching of the particles.}
\label{fig:magnetization}
\end{figure}

Finally, we determine the resulting particle center-to-center distances $r_{12}$ for each considered magnitude $B^\text{ext}$ of the external magnetic field as a function of the history of the system. To this end, we identify the intersections between the curves for $F^\text{mag}(r_{12})$ (spline curves) and $F^\text{el}(r_{12})$ (rod model), see Fig.~\ref{fig:Force}. As %Fig.~\ref{fig:Force} 
the figure also illustrates, there are three qualitatively different situations: %the curves of the magnetic attraction force can be divided in the general three types:
%JOURNALE (ZUMINDEST UNSERE) MOEGEN SOLCHE LISTEN IM TEXT NICHT...
%\begin{itemize}
%  \item[a)] 
(i) only one intersection between $F^\text{el}(r_{12})$ and $F^\text{mag}(r_{12})$ close to $r_{12}/a \approx 3.5$, see the case of $B^\text{ext}=\SI{13}{\milli\tesla}$;
%   \item[b)] 
(ii) three intersections, one located at $r_{12}/a \approx 3.5$, one at \mbox{$r_{12}/a \approx 2$}, and one between these values, see the case of $B^\text{ext}=\SI{39}{\milli\tesla}$;
%   \item[c)] 
and (iii) only one intersection close to  $r_{12}/a \approx 2$, see the case of $B^\text{ext}=\SI{65}{\milli\tesla}$.
%\end{itemize}

Increasing the magnetic field from \mbox{$\SI{0}{\milli\tesla}$}, our evaluations demonstrate that there are states of balanced magnetic and mechanical forces and significant particle separation for center-to-center distances down to $r_{12}/a \approx 2.95$. Initially, this corresponds to the intersection described by case (i), then to the first intersection of case (ii). The intermediate intersection of case (ii) corresponds to a metastable saddle configuration in the overall energy, see Fig.~\ref{fig5}(c). At an external field strength of $B^\text{ext}\approx\SI{63.05}{mT}$, the simulation predicts a switch from case (ii) to case (iii). Thus the state of separated particles collapses towards a state of virtual touching of $r_{12}/a \approx 2$. %If $B^\text{ext}$ is now increased up to $\SI{65}{\milli\tesla}$ and subsequently decreased, 
Subsequently decreasing $B^\text{ext}$, the system switches from case (iii) back to case (ii). However, the particles now remain in the state of virtual touching of $r_{12}/a \approx 2$ corresponding to the third intersection of case (ii). Only at $B^\text{ext} \approx \SI{30.55}{\milli\tesla}$, the set-up switches back to case (i), which implies that the particles re-separate. Consequently, the FE simulations confirm the scenario of hysteresis predicted qualitatively by the theoretical analysis in Secs.~\ref{linear-elastic} and \ref{dipole-spring}. The corresponding simulation results %from the FE simulations 
are shown by the filled circles in Fig.~\ref{fig3}. 

More precisely, the FE simulations provide the following additional and more quantitative insights. 
As Fig.~\ref{fig3} demonstrates, the theoretical analyses in Secs.~\ref{linear-elastic} and \ref{dipole-spring} describe the system in the non-collapsed state quantitatively correctly in a broad initial interval of increasing amplitude of the magnetic field. Thus, for well-separated spherical particles in an elastic matrix, the theoretical schemes developed in Refs.~\citenum{kim1994faxen,phan1994load,puljiz2016forces, puljiz2017forces,menzel2017force} provide an efficient and accurate characterization. 

The point of collapse is predicted correctly by the analytical theory to good approximation. Our FE simulations indicate a slightly elevated amplitude of the magnetic field at which the collapse occurs. %This effect is not simply explained by nonlinear elastic contributions. 
We thus have repeated the simulations, now following a linearized law of magnetization as in Eq.~(\ref{magnetization_1}) instead of the nonlinear relation of Eq.~(\ref{eq:Langevin}). The linearized law allows for an in principle unbounded growth of the magnetization, resulting in stronger magnetic attraction. Consequently, the corresponding simulations predict an earlier collapse of the separated state in quite good agreement with the theoretical results, see the triangles in Fig.~\ref{fig3}. However, the situation is more complex than might be expected from this agreement of the data. % the theory is not simply adjusted by introducing the nonlinear magnetization law instead. 
In the simulations, which resolve the spatial inhomogeneity of the magnetization across the particles, see below, the linearization mainly affects those parts of the particles that are closest to each other and thus are most strongly magnetized. In contrast to that, in our theory the dipoles are concentrated in the more distanced particle centers and are not as severely affected by the linearization. 

In the collapsed state, the extrapolation of the simulation curves predict that a complete touching of the particles does not occur. In such a situation, an extreme compression of the elastic material between the particles would be necessary, if the assumed no-slip anchoring of the polymeric matrix on the surfaces of the particles persists. This leads to values of $r_{12}/a$ slightly larger than $2$ as stressed by the inset in Fig.~\ref{fig3}. Our experimental resolution did not allow us to further clarify this issue on the actual experimental system. 

Most importantly, the FE simulations reveal an even more pronounced hysteretic behavior. The magnitude of the external magnetic field for re-separation of the particles is found to be significantly lower than predicted from the theoretical analyses in Secs.~\ref{linear-elastic} and \ref{dipole-spring}, see Fig.~\ref{fig3}. Particularly, this is due to the approximation in terms of magnetic dipoles located at the particle centers. % and the employed magnetization law. 
Our FE simulations reveal that, in the state of virtual touching, the magnetization within the particles is strongly inhomogeneous. Around the virtual contact points of the nearly touching particles, local magnetic field amplitudes of up to $\SI{0.66}{\tesla}$ are found inside the particles, despite the relatively small value of the maximum external magnetic field amplitude of $\SI{65}{\milli\tesla}$. %, it was shown by the FE simulations that the local magnetic field $\mathbf B^\text{in}$ reaches values up to $\SI{0.66}{\tesla}$ if the particles are close to each other. 
In this case, a spatial resolution of the magnetization field inside the particles becomes important, together with the nonlinear magnetization behavior according to Eq.~(\ref{eq:Langevin}) in the vicinity of the virtual contact points between the particles. % to calculate the magnetic attraction force in a correct way.
The local magnitude of the magnetization within quarters of the particles is illustrated in Fig.~\ref{fig:magnetization} for a separated state at $B^\text{ext}=\SI{63.05}{\milli\tesla}$ and a state of virtual touching at $B^\text{ext}=\SI{65.00}{\milli\tesla}$. % of the system if the field is increased starting from $\SI{0}{\milli\tesla}$. 
Naturally, not bounding the magnetization by using a linearized magnetization law allows for even stronger magnetization in the near-touching parts of the particles. In such a case, corresponding simulation results predict an even more pronounced hysteresis, see the triangles in Fig.~\ref{fig3}. 

%For the theoretical modeling in Secs.~\ref{linear-elastic} and \ref{dipole-spring} the dipoles were centered in the magnetic particles so that they are separated by at least one particle diameter.
Nevertheless, from Fig.~\ref{fig3} we note that the different ways of description all lead to results that agree well with the experimental data within the experimental error bars.
A higher resolution of measurement will allow to further develop and specify the theoretical and numerical tools in the future.

%The described simulations confirm the prediction of hysteresis loops according to Sec.~\ref{linear-elastic} and \ref{dipole-spring}. They are able to improve the analytical solution by quantitative corrections in the close to the touching state of the particles which result from the the dipole-dipole approximation and the linear elastic approximation in the analytical theories. As it becomes apparent from Figs.~\ref{fig:Force} and \ref{fig:magnetization}, these approximations do not hold in the collapsed state. 

\section{Conclusions}
\label{conclusion}

In summary, we have experimentally observed the reversible approach and separation of two paramagnetic metallic particles in a soft elastic gel matrix, induced by adjusting the magnitude of an external magnetic field. Above a certain threshold of the magnetic field, the particles collapsed into a virtually touching state. We have analyzed the scenario theoretically using a dipolar approximation for the magnetic interactions and linear elasticity theory to describe the distortions of the elastic matrix. This description was further simplified by projecting it onto reduced dipole-spring models. Significant hysteretic behavior was revealed in this way. That is, the collapse into the state of virtual touching occurred at a significantly larger magnetic field than their sudden separation back into a well-distanced state when subsequently decreasing the field amplitude. Finally, we further quantified the situation by additional finite-element simulations. They spatially resolve the magnetic field also inside the magnetic particles and allow for nonlinear elasticity of the elastic environment. Our simulations are in agreement with the experimental observations and to good approximation confirm our theoretical analysis until the collapse into the virtually touching state occurs. However, %As can be understood from their more detailed calculation of the magnetic field, 
they predict a more pronounced hysteresis with the separation of the particles occurring at a lower magnetic field amplitude than calculated from the linearized analytical theory. 

We expect our results, which explicitly demonstrate the reversible externally induced virtual touching and separation of hard particles in a soft elastic matrix, to be of high practical relevance from an application point of view. A very illustrative example is certainly a switchable damping device. Instead of using only two particles, one may arrange many particles in parallel rows. Then, in the absence of magnetic fields, the material of separated particles is soft under compression along the axis of the rows. When the particles enter a state of virtual touching under magnetic fields, the resulting aggregates significantly harden \cite{annunziata2013hardening}.

In a broader framework, our investigations are related to various studies in several other areas. 
Far away from the two particles, the distortion induced in the elastic matrix resembles that of a point-like mechanical force dipole. 
In such an approximation, for example, the stress exerted by active biological cells on their environment has been addressed \cite{schwarz2002elastic,schwarz2013physics}. For instance, preferred mutual orientations were explained via induced long-ranged elasticity-mediated interactions between the cells \cite{bischofs2004elastic,yuval2013dynamics}. Similarly, localized force dipoles and their mutual interactions by distortion of their elastic environment are treated in the theory of defects in crystal structures \cite{teodosiu1982elastic}. In a different context, the question of whether the separated state of the two magnetic particles is stable, and at which point this state collapses and the particles touch each other, has recently been studied in the context of magnetosome filaments \cite{boltz2017buckling}. These elastic elements are found, for instance, in magnetotactic bacteria that detect the magnetic field of the earth for their orientation. Elastic filaments connecting the magnetic particles need to provide sufficient rigidity to avoid the particle collapse. 

We believe that our results are important for the future construction of tunable dampers and vibration absorbers, soft actuators, and energy storage devices from elastic composite materials. Here, we mainly concentrated on magnetically induced effects. Yet, many of the described properties may carry over to the case of electric fields \cite{allahyarov2015simulation,allahyarov2016dipole}. There, also the question of energy storage will be more prevalent \cite{li2007electric,kim2009high,li2009nanocomposites, wang2011polymer,fredin2013substantial,allahyarov2016dipole}. In that case, however, touching of the particles together with the formation of chained structures should be hindered to avoid electric short circuits.

\begin{acknowledgments}
The authors thank the Deutsche Forschungsgemeinschaft for support of this work through the priority program SPP 1681, grant nos.\ OD 18/21 (JN, SO), KA 3309/2 (KK, MK), AU 321/3 (SH, GKA), and ME 3571/3 (MP, AMM). 
\end{acknowledgments}

%%%Bibliography

%\bibliography{lit_touching_2017nov12}

\begin{thebibliography}{104}
\expandafter\ifx\csname natexlab\endcsname\relax\def\natexlab#1{#1}\fi
\expandafter\ifx\csname bibnamefont\endcsname\relax
  \def\bibnamefont#1{#1}\fi
\expandafter\ifx\csname bibfnamefont\endcsname\relax
  \def\bibfnamefont#1{#1}\fi
\expandafter\ifx\csname citenamefont\endcsname\relax
  \def\citenamefont#1{#1}\fi
\expandafter\ifx\csname url\endcsname\relax
  \def\url#1{\texttt{#1}}\fi
\expandafter\ifx\csname urlprefix\endcsname\relax\def\urlprefix{URL }\fi
\providecommand{\bibinfo}[2]{#2}
\providecommand{\eprint}[2][]{\url{#2}}

\bibitem[{\citenamefont{An and Shaw}(2003)}]{an2003actuating}
\bibinfo{author}{\bibfnamefont{Y.}~\bibnamefont{An}} \bibnamefont{and}
  \bibinfo{author}{\bibfnamefont{M.~T.} \bibnamefont{Shaw}},
  \bibinfo{journal}{Smart Mater. Struct.} \textbf{\bibinfo{volume}{12}},
  \bibinfo{pages}{157} (\bibinfo{year}{2003}).

\bibitem[{\citenamefont{Filipcsei et~al.}(2007)\citenamefont{Filipcsei,
  Csetneki, Szil{\'a}gyi, and Zr\'{i}nyi}}]{filipcsei2007magnetic}
\bibinfo{author}{\bibfnamefont{G.}~\bibnamefont{Filipcsei}},
  \bibinfo{author}{\bibfnamefont{I.}~\bibnamefont{Csetneki}},
  \bibinfo{author}{\bibfnamefont{A.}~\bibnamefont{Szil{\'a}gyi}},
  \bibnamefont{and}
  \bibinfo{author}{\bibfnamefont{M.}~\bibnamefont{Zr\'{i}nyi}},
  \bibinfo{journal}{Adv. Polym. Sci.} \textbf{\bibinfo{volume}{206}},
  \bibinfo{pages}{137} (\bibinfo{year}{2007}).

\bibitem[{\citenamefont{Zimmermann et~al.}(2007)\citenamefont{Zimmermann,
  Naletova, Zeidis, Turkov, Kolev, Lukashevich, and
  Stepanov}}]{zimmermann2007deformable}
\bibinfo{author}{\bibfnamefont{K.}~\bibnamefont{Zimmermann}},
  \bibinfo{author}{\bibfnamefont{V.~A.} \bibnamefont{Naletova}},
  \bibinfo{author}{\bibfnamefont{I.}~\bibnamefont{Zeidis}},
  \bibinfo{author}{\bibfnamefont{V.~A.} \bibnamefont{Turkov}},
  \bibinfo{author}{\bibfnamefont{E.}~\bibnamefont{Kolev}},
  \bibinfo{author}{\bibfnamefont{M.~V.} \bibnamefont{Lukashevich}},
  \bibnamefont{and} \bibinfo{author}{\bibfnamefont{G.~V.}
  \bibnamefont{Stepanov}}, \bibinfo{journal}{J. Magn. Magn. Mater.}
  \textbf{\bibinfo{volume}{311}}, \bibinfo{pages}{450} (\bibinfo{year}{2007}).

\bibitem[{\citenamefont{Raikher et~al.}(2008)\citenamefont{Raikher, Stolbov,
  and Stepanov}}]{raikher2008shape}
\bibinfo{author}{\bibfnamefont{Y.~L.} \bibnamefont{Raikher}},
  \bibinfo{author}{\bibfnamefont{O.~V.} \bibnamefont{Stolbov}},
  \bibnamefont{and} \bibinfo{author}{\bibfnamefont{G.~V.}
  \bibnamefont{Stepanov}}, \bibinfo{journal}{J. Phys. D: Appl. Phys.}
  \textbf{\bibinfo{volume}{41}}, \bibinfo{pages}{152002}
  (\bibinfo{year}{2008}).

\bibitem[{\citenamefont{Fuhrer et~al.}(2009)\citenamefont{Fuhrer, Athanassiou,
  Luechinger, and Stark}}]{fuhrer2009crosslinking}
\bibinfo{author}{\bibfnamefont{R.}~\bibnamefont{Fuhrer}},
  \bibinfo{author}{\bibfnamefont{E.~K.} \bibnamefont{Athanassiou}},
  \bibinfo{author}{\bibfnamefont{N.~A.} \bibnamefont{Luechinger}},
  \bibnamefont{and} \bibinfo{author}{\bibfnamefont{W.~J.} \bibnamefont{Stark}},
  \bibinfo{journal}{Small} \textbf{\bibinfo{volume}{5}}, \bibinfo{pages}{383}
  (\bibinfo{year}{2009}).

\bibitem[{\citenamefont{B{\"o}se et~al.}(2012)\citenamefont{B{\"o}se,
  Rabindranath, and Ehrlich}}]{bose2012soft}
\bibinfo{author}{\bibfnamefont{H.}~\bibnamefont{B{\"o}se}},
  \bibinfo{author}{\bibfnamefont{R.}~\bibnamefont{Rabindranath}},
  \bibnamefont{and} \bibinfo{author}{\bibfnamefont{J.}~\bibnamefont{Ehrlich}},
  \bibinfo{journal}{J. Intel. Mater. Syst. Struct.}
  \textbf{\bibinfo{volume}{23}}, \bibinfo{pages}{989} (\bibinfo{year}{2012}).

\bibitem[{\citenamefont{Ilg}(2013)}]{ilg2013stimuli}
\bibinfo{author}{\bibfnamefont{P.}~\bibnamefont{Ilg}}, \bibinfo{journal}{Soft
  Matter} \textbf{\bibinfo{volume}{9}}, \bibinfo{pages}{3465}
  (\bibinfo{year}{2013}).

\bibitem[{\citenamefont{Deng et~al.}(2006)\citenamefont{Deng, Gong, and
  Wang}}]{deng2006development}
\bibinfo{author}{\bibfnamefont{H.-x.} \bibnamefont{Deng}},
  \bibinfo{author}{\bibfnamefont{X.-l.} \bibnamefont{Gong}}, \bibnamefont{and}
  \bibinfo{author}{\bibfnamefont{L.-h.} \bibnamefont{Wang}},
  \bibinfo{journal}{Smart Mater. Struct.} \textbf{\bibinfo{volume}{15}},
  \bibinfo{pages}{N111} (\bibinfo{year}{2006}).

\bibitem[{\citenamefont{Sun et~al.}(2008)\citenamefont{Sun, Gong, Jiang, Li,
  Xu, and Li}}]{sun2008study}
\bibinfo{author}{\bibfnamefont{T.~L.} \bibnamefont{Sun}},
  \bibinfo{author}{\bibfnamefont{X.~L.} \bibnamefont{Gong}},
  \bibinfo{author}{\bibfnamefont{W.~Q.} \bibnamefont{Jiang}},
  \bibinfo{author}{\bibfnamefont{J.~F.} \bibnamefont{Li}},
  \bibinfo{author}{\bibfnamefont{Z.~B.} \bibnamefont{Xu}}, \bibnamefont{and}
  \bibinfo{author}{\bibfnamefont{W.~H.} \bibnamefont{Li}},
  \bibinfo{journal}{Polym. Test.} \textbf{\bibinfo{volume}{27}},
  \bibinfo{pages}{520} (\bibinfo{year}{2008}).

\bibitem[{\citenamefont{Liao et~al.}(2012)\citenamefont{Liao, Gong, Xuan, Kang,
  and Zong}}]{liao2012development}
\bibinfo{author}{\bibfnamefont{G.~J.} \bibnamefont{Liao}},
  \bibinfo{author}{\bibfnamefont{X.~L.} \bibnamefont{Gong}},
  \bibinfo{author}{\bibfnamefont{S.~H.} \bibnamefont{Xuan}},
  \bibinfo{author}{\bibfnamefont{C.~J.} \bibnamefont{Kang}}, \bibnamefont{and}
  \bibinfo{author}{\bibfnamefont{L.~H.} \bibnamefont{Zong}},
  \bibinfo{journal}{J. Int. Mater. Syst. Struct.}
  \textbf{\bibinfo{volume}{23}}, \bibinfo{pages}{25} (\bibinfo{year}{2012}).

\bibitem[{\citenamefont{Molchanov et~al.}(2014)\citenamefont{Molchanov,
  Stepanov, Vasiliev, Kramarenko, Khokhlov, Xu, and
  Guo}}]{molchanov2014viscoelastic}
\bibinfo{author}{\bibfnamefont{V.~S.} \bibnamefont{Molchanov}},
  \bibinfo{author}{\bibfnamefont{G.~V.} \bibnamefont{Stepanov}},
  \bibinfo{author}{\bibfnamefont{V.~G.} \bibnamefont{Vasiliev}},
  \bibinfo{author}{\bibfnamefont{E.~Y.} \bibnamefont{Kramarenko}},
  \bibinfo{author}{\bibfnamefont{A.~R.} \bibnamefont{Khokhlov}},
  \bibinfo{author}{\bibfnamefont{Z.-D.} \bibnamefont{Xu}}, \bibnamefont{and}
  \bibinfo{author}{\bibfnamefont{Y.-Q.} \bibnamefont{Guo}},
  \bibinfo{journal}{Macromol. Mater. Eng.} \textbf{\bibinfo{volume}{299}},
  \bibinfo{pages}{1116} (\bibinfo{year}{2014}).

\bibitem[{\citenamefont{Mietta et~al.}(2016)\citenamefont{Mietta, Tamborenea,
  and Negri}}]{mietta2016anisotropic}
\bibinfo{author}{\bibfnamefont{J.~L.} \bibnamefont{Mietta}},
  \bibinfo{author}{\bibfnamefont{P.~I.} \bibnamefont{Tamborenea}},
  \bibnamefont{and} \bibinfo{author}{\bibfnamefont{R.~M.} \bibnamefont{Negri}},
  \bibinfo{journal}{Soft Matter} \textbf{\bibinfo{volume}{12}},
  \bibinfo{pages}{6430} (\bibinfo{year}{2016}).

\bibitem[{\citenamefont{Li et~al.}(2007)\citenamefont{Li, Zhang, and
  Ducharme}}]{li2007electric}
\bibinfo{author}{\bibfnamefont{J.~Y.} \bibnamefont{Li}},
  \bibinfo{author}{\bibfnamefont{L.}~\bibnamefont{Zhang}}, \bibnamefont{and}
  \bibinfo{author}{\bibfnamefont{S.}~\bibnamefont{Ducharme}},
  \bibinfo{journal}{Appl. Phys. Lett.} \textbf{\bibinfo{volume}{90}},
  \bibinfo{pages}{132901} (\bibinfo{year}{2007}).

\bibitem[{\citenamefont{Kim et~al.}(2009)\citenamefont{Kim, Doss, Tillotson,
  Hotchkiss, Pan, Marder, Li, Calame, and Perry}}]{kim2009high}
\bibinfo{author}{\bibfnamefont{P.}~\bibnamefont{Kim}},
  \bibinfo{author}{\bibfnamefont{N.~M.} \bibnamefont{Doss}},
  \bibinfo{author}{\bibfnamefont{J.~P.} \bibnamefont{Tillotson}},
  \bibinfo{author}{\bibfnamefont{P.~J.} \bibnamefont{Hotchkiss}},
  \bibinfo{author}{\bibfnamefont{M.-J.} \bibnamefont{Pan}},
  \bibinfo{author}{\bibfnamefont{S.~R.} \bibnamefont{Marder}},
  \bibinfo{author}{\bibfnamefont{J.}~\bibnamefont{Li}},
  \bibinfo{author}{\bibfnamefont{J.~P.} \bibnamefont{Calame}},
  \bibnamefont{and} \bibinfo{author}{\bibfnamefont{J.~W.} \bibnamefont{Perry}},
  \bibinfo{journal}{ACS Nano} \textbf{\bibinfo{volume}{3}},
  \bibinfo{pages}{2581} (\bibinfo{year}{2009}).

\bibitem[{\citenamefont{Li et~al.}(2009)\citenamefont{Li, Seok, Chu, Dogan,
  Zhang, and Wang}}]{li2009nanocomposites}
\bibinfo{author}{\bibfnamefont{J.}~\bibnamefont{Li}},
  \bibinfo{author}{\bibfnamefont{S.~I.} \bibnamefont{Seok}},
  \bibinfo{author}{\bibfnamefont{B.}~\bibnamefont{Chu}},
  \bibinfo{author}{\bibfnamefont{F.}~\bibnamefont{Dogan}},
  \bibinfo{author}{\bibfnamefont{Q.}~\bibnamefont{Zhang}}, \bibnamefont{and}
  \bibinfo{author}{\bibfnamefont{Q.}~\bibnamefont{Wang}},
  \bibinfo{journal}{Adv. Mater.} \textbf{\bibinfo{volume}{21}},
  \bibinfo{pages}{217} (\bibinfo{year}{2009}).

\bibitem[{\citenamefont{Wang and Zhu}(2011)}]{wang2011polymer}
\bibinfo{author}{\bibfnamefont{Q.}~\bibnamefont{Wang}} \bibnamefont{and}
  \bibinfo{author}{\bibfnamefont{L.}~\bibnamefont{Zhu}}, \bibinfo{journal}{J.
  Polym. Sci. B: Polym. Phys.} \textbf{\bibinfo{volume}{49}},
  \bibinfo{pages}{1421} (\bibinfo{year}{2011}).

\bibitem[{\citenamefont{Fredin et~al.}(2013)\citenamefont{Fredin, Li, Lanagan,
  Ratner, and Marks}}]{fredin2013substantial}
\bibinfo{author}{\bibfnamefont{L.~A.} \bibnamefont{Fredin}},
  \bibinfo{author}{\bibfnamefont{Z.}~\bibnamefont{Li}},
  \bibinfo{author}{\bibfnamefont{M.~T.} \bibnamefont{Lanagan}},
  \bibinfo{author}{\bibfnamefont{M.~A.} \bibnamefont{Ratner}},
  \bibnamefont{and} \bibinfo{author}{\bibfnamefont{T.~J.} \bibnamefont{Marks}},
  \bibinfo{journal}{Adv. Func. Mater.} \textbf{\bibinfo{volume}{23}},
  \bibinfo{pages}{3560} (\bibinfo{year}{2013}).

\bibitem[{\citenamefont{Allahyarov et~al.}(2016)\citenamefont{Allahyarov,
  L{\"o}wen, and Zhu}}]{allahyarov2016dipole}
\bibinfo{author}{\bibfnamefont{E.}~\bibnamefont{Allahyarov}},
  \bibinfo{author}{\bibfnamefont{H.}~\bibnamefont{L{\"o}wen}},
  \bibnamefont{and} \bibinfo{author}{\bibfnamefont{L.}~\bibnamefont{Zhu}},
  \bibinfo{journal}{Phys. Chem. Chem. Phys.} \textbf{\bibinfo{volume}{18}},
  \bibinfo{pages}{19103} (\bibinfo{year}{2016}).

\bibitem[{\citenamefont{Zr{\'\i}nyi et~al.}(1996)\citenamefont{Zr{\'\i}nyi,
  Barsi, and B{\"u}ki}}]{zrinyi1996deformation}
\bibinfo{author}{\bibfnamefont{M.}~\bibnamefont{Zr{\'\i}nyi}},
  \bibinfo{author}{\bibfnamefont{L.}~\bibnamefont{Barsi}}, \bibnamefont{and}
  \bibinfo{author}{\bibfnamefont{A.}~\bibnamefont{B{\"u}ki}},
  \bibinfo{journal}{J. Chem. Phys.} \textbf{\bibinfo{volume}{104}},
  \bibinfo{pages}{8750} (\bibinfo{year}{1996}).

\bibitem[{\citenamefont{Collin et~al.}(2003)\citenamefont{Collin, Auernhammer,
  Gavat, Martinoty, and Brand}}]{collin2003frozen}
\bibinfo{author}{\bibfnamefont{D.}~\bibnamefont{Collin}},
  \bibinfo{author}{\bibfnamefont{G.~K.} \bibnamefont{Auernhammer}},
  \bibinfo{author}{\bibfnamefont{O.}~\bibnamefont{Gavat}},
  \bibinfo{author}{\bibfnamefont{P.}~\bibnamefont{Martinoty}},
  \bibnamefont{and} \bibinfo{author}{\bibfnamefont{H.~R.} \bibnamefont{Brand}},
  \bibinfo{journal}{Macromol. Rapid Commun.} \textbf{\bibinfo{volume}{24}},
  \bibinfo{pages}{737} (\bibinfo{year}{2003}).

\bibitem[{\citenamefont{Jackson}(1999)}]{jackson1962classical}
\bibinfo{author}{\bibfnamefont{J.~D.} \bibnamefont{Jackson}},
  \emph{\bibinfo{title}{Classical Electrodynamics}} (\bibinfo{publisher}{Wiley,
  New York}, \bibinfo{year}{1999}).

\bibitem[{\citenamefont{Diguet et~al.}(2009)\citenamefont{Diguet, Beaugnon, and
  Cavaill{\'e}}}]{diguet2009dipolar}
\bibinfo{author}{\bibfnamefont{G.}~\bibnamefont{Diguet}},
  \bibinfo{author}{\bibfnamefont{E.}~\bibnamefont{Beaugnon}}, \bibnamefont{and}
  \bibinfo{author}{\bibfnamefont{J.-Y.} \bibnamefont{Cavaill{\'e}}},
  \bibinfo{journal}{J. Magn. Magn. Mater.} \textbf{\bibinfo{volume}{321}},
  \bibinfo{pages}{396} (\bibinfo{year}{2009}).

\bibitem[{\citenamefont{Stolbov et~al.}(2011)\citenamefont{Stolbov, Raikher,
  and Balasoiu}}]{stolbov2011modelling}
\bibinfo{author}{\bibfnamefont{O.~V.} \bibnamefont{Stolbov}},
  \bibinfo{author}{\bibfnamefont{Y.~L.} \bibnamefont{Raikher}},
  \bibnamefont{and} \bibinfo{author}{\bibfnamefont{M.}~\bibnamefont{Balasoiu}},
  \bibinfo{journal}{Soft Matter} \textbf{\bibinfo{volume}{7}},
  \bibinfo{pages}{8484} (\bibinfo{year}{2011}).

\bibitem[{\citenamefont{Ivaneyko et~al.}(2012)\citenamefont{Ivaneyko,
  Toshchevikov, Saphiannikova, and Heinrich}}]{ivaneyko2012effects}
\bibinfo{author}{\bibfnamefont{D.}~\bibnamefont{Ivaneyko}},
  \bibinfo{author}{\bibfnamefont{V.}~\bibnamefont{Toshchevikov}},
  \bibinfo{author}{\bibfnamefont{M.}~\bibnamefont{Saphiannikova}},
  \bibnamefont{and} \bibinfo{author}{\bibfnamefont{G.}~\bibnamefont{Heinrich}},
  \bibinfo{journal}{Condens. Matter Phys.} \textbf{\bibinfo{volume}{15}},
  \bibinfo{pages}{33601} (\bibinfo{year}{2012}).

\bibitem[{\citenamefont{Zubarev}(2013{\natexlab{a}})}]{zubarev2013effect}
\bibinfo{author}{\bibfnamefont{A.~Y.} \bibnamefont{Zubarev}},
  \bibinfo{journal}{Soft Matter} \textbf{\bibinfo{volume}{9}},
  \bibinfo{pages}{4985} (\bibinfo{year}{2013}{\natexlab{a}}).

\bibitem[{\citenamefont{Menzel}(2015)}]{menzel2015tuned}
\bibinfo{author}{\bibfnamefont{A.~M.} \bibnamefont{Menzel}},
  \bibinfo{journal}{Phys. Rep.} \textbf{\bibinfo{volume}{554}},
  \bibinfo{pages}{1} (\bibinfo{year}{2015}).

\bibitem[{\citenamefont{Allahyarov et~al.}(2015)\citenamefont{Allahyarov,
  L{\"o}wen, and Zhu}}]{allahyarov2015simulation}
\bibinfo{author}{\bibfnamefont{E.}~\bibnamefont{Allahyarov}},
  \bibinfo{author}{\bibfnamefont{H.}~\bibnamefont{L{\"o}wen}},
  \bibnamefont{and} \bibinfo{author}{\bibfnamefont{L.}~\bibnamefont{Zhu}},
  \bibinfo{journal}{Phys. Chem. Chem. Phys.} \textbf{\bibinfo{volume}{17}},
  \bibinfo{pages}{32479} (\bibinfo{year}{2015}).

\bibitem[{\citenamefont{Huang et~al.}(2016)\citenamefont{Huang, Pessot, Cremer,
  Weeber, Holm, Nowak, Odenbach, Menzel, and Auernhammer}}]{huang2016buckling}
\bibinfo{author}{\bibfnamefont{S.}~\bibnamefont{Huang}},
  \bibinfo{author}{\bibfnamefont{G.}~\bibnamefont{Pessot}},
  \bibinfo{author}{\bibfnamefont{P.}~\bibnamefont{Cremer}},
  \bibinfo{author}{\bibfnamefont{R.}~\bibnamefont{Weeber}},
  \bibinfo{author}{\bibfnamefont{C.}~\bibnamefont{Holm}},
  \bibinfo{author}{\bibfnamefont{J.}~\bibnamefont{Nowak}},
  \bibinfo{author}{\bibfnamefont{S.}~\bibnamefont{Odenbach}},
  \bibinfo{author}{\bibfnamefont{A.~M.} \bibnamefont{Menzel}},
  \bibnamefont{and} \bibinfo{author}{\bibfnamefont{G.~K.}
  \bibnamefont{Auernhammer}}, \bibinfo{journal}{Soft Matter}
  \textbf{\bibinfo{volume}{12}}, \bibinfo{pages}{228} (\bibinfo{year}{2016}).

\bibitem[{\citenamefont{Romeis et~al.}(2016)\citenamefont{Romeis, Toshchevikov,
  and Saphiannikova}}]{romeis2016elongated}
\bibinfo{author}{\bibfnamefont{D.}~\bibnamefont{Romeis}},
  \bibinfo{author}{\bibfnamefont{V.}~\bibnamefont{Toshchevikov}},
  \bibnamefont{and}
  \bibinfo{author}{\bibfnamefont{M.}~\bibnamefont{Saphiannikova}},
  \bibinfo{journal}{Soft Matter} \textbf{\bibinfo{volume}{12}},
  \bibinfo{pages}{9364} (\bibinfo{year}{2016}).

\bibitem[{\citenamefont{Metsch et~al.}(2016)\citenamefont{Metsch, Kalina,
  Spieler, and K{\"a}stner}}]{metsch2016numerical}
\bibinfo{author}{\bibfnamefont{P.}~\bibnamefont{Metsch}},
  \bibinfo{author}{\bibfnamefont{K.~A.} \bibnamefont{Kalina}},
  \bibinfo{author}{\bibfnamefont{C.}~\bibnamefont{Spieler}}, \bibnamefont{and}
  \bibinfo{author}{\bibfnamefont{M.}~\bibnamefont{K{\"a}stner}},
  \bibinfo{journal}{Comp. Mater. Sci.} \textbf{\bibinfo{volume}{124}},
  \bibinfo{pages}{364} (\bibinfo{year}{2016}).

\bibitem[{\citenamefont{Jolly et~al.}(1996{\natexlab{a}})\citenamefont{Jolly,
  Carlson, Mu{\~n}oz, and Bullions}}]{jolly1996magnetoviscoelastic}
\bibinfo{author}{\bibfnamefont{M.~R.} \bibnamefont{Jolly}},
  \bibinfo{author}{\bibfnamefont{J.~D.} \bibnamefont{Carlson}},
  \bibinfo{author}{\bibfnamefont{B.~C.} \bibnamefont{Mu{\~n}oz}},
  \bibnamefont{and} \bibinfo{author}{\bibfnamefont{T.~A.}
  \bibnamefont{Bullions}}, \bibinfo{journal}{J. Intel. Mater. Syst. Struct.}
  \textbf{\bibinfo{volume}{7}}, \bibinfo{pages}{613}
  (\bibinfo{year}{1996}{\natexlab{a}}).

\bibitem[{\citenamefont{Jolly et~al.}(1996{\natexlab{b}})\citenamefont{Jolly,
  Carlson, and Mu{\~n}oz}}]{jolly1996model}
\bibinfo{author}{\bibfnamefont{M.~R.} \bibnamefont{Jolly}},
  \bibinfo{author}{\bibfnamefont{J.~D.} \bibnamefont{Carlson}},
  \bibnamefont{and} \bibinfo{author}{\bibfnamefont{B.~C.}
  \bibnamefont{Mu{\~n}oz}}, \bibinfo{journal}{Smart Mater. Struct.}
  \textbf{\bibinfo{volume}{5}}, \bibinfo{pages}{607}
  (\bibinfo{year}{1996}{\natexlab{b}}).

\bibitem[{\citenamefont{Jarkova et~al.}(2003)\citenamefont{Jarkova, Pleiner,
  M{\"u}ller, and Brand}}]{jarkova2003hydrodynamics}
\bibinfo{author}{\bibfnamefont{E.}~\bibnamefont{Jarkova}},
  \bibinfo{author}{\bibfnamefont{H.}~\bibnamefont{Pleiner}},
  \bibinfo{author}{\bibfnamefont{H.-W.} \bibnamefont{M{\"u}ller}},
  \bibnamefont{and} \bibinfo{author}{\bibfnamefont{H.~R.} \bibnamefont{Brand}},
  \bibinfo{journal}{Phys. Rev. E} \textbf{\bibinfo{volume}{68}},
  \bibinfo{pages}{041706} (\bibinfo{year}{2003}).

\bibitem[{\citenamefont{Stepanov et~al.}(2007)\citenamefont{Stepanov,
  Abramchuk, Grishin, Nikitin, Kramarenko, and Khokhlov}}]{stepanov2007effect}
\bibinfo{author}{\bibfnamefont{G.~V.} \bibnamefont{Stepanov}},
  \bibinfo{author}{\bibfnamefont{S.~S.} \bibnamefont{Abramchuk}},
  \bibinfo{author}{\bibfnamefont{D.~A.} \bibnamefont{Grishin}},
  \bibinfo{author}{\bibfnamefont{L.~V.} \bibnamefont{Nikitin}},
  \bibinfo{author}{\bibfnamefont{E.~Y.} \bibnamefont{Kramarenko}},
  \bibnamefont{and} \bibinfo{author}{\bibfnamefont{A.~R.}
  \bibnamefont{Khokhlov}}, \bibinfo{journal}{Polymer}
  \textbf{\bibinfo{volume}{48}}, \bibinfo{pages}{488} (\bibinfo{year}{2007}).

\bibitem[{\citenamefont{B{\"o}se and
  R{\"o}der}(2009)}]{bose2009magnetorheological}
\bibinfo{author}{\bibfnamefont{H.}~\bibnamefont{B{\"o}se}} \bibnamefont{and}
  \bibinfo{author}{\bibfnamefont{R.}~\bibnamefont{R{\"o}der}},
  \bibinfo{journal}{J. Phys.: Conf. Ser.} \textbf{\bibinfo{volume}{149}},
  \bibinfo{pages}{012090} (\bibinfo{year}{2009}).

\bibitem[{\citenamefont{Chertovich et~al.}(2010)\citenamefont{Chertovich,
  Stepanov, Kramarenko, and Khokhlov}}]{chertovich2010new}
\bibinfo{author}{\bibfnamefont{A.~V.} \bibnamefont{Chertovich}},
  \bibinfo{author}{\bibfnamefont{G.~V.} \bibnamefont{Stepanov}},
  \bibinfo{author}{\bibfnamefont{E.~Y.} \bibnamefont{Kramarenko}},
  \bibnamefont{and} \bibinfo{author}{\bibfnamefont{A.~R.}
  \bibnamefont{Khokhlov}}, \bibinfo{journal}{Macromolecul. Mater. Eng.}
  \textbf{\bibinfo{volume}{295}}, \bibinfo{pages}{336} (\bibinfo{year}{2010}).

\bibitem[{\citenamefont{Wood and Camp}(2011)}]{wood2011modeling}
\bibinfo{author}{\bibfnamefont{D.~S.} \bibnamefont{Wood}} \bibnamefont{and}
  \bibinfo{author}{\bibfnamefont{P.~J.} \bibnamefont{Camp}},
  \bibinfo{journal}{Phys. Rev. E} \textbf{\bibinfo{volume}{83}},
  \bibinfo{pages}{011402} (\bibinfo{year}{2011}).

\bibitem[{\citenamefont{Evans et~al.}(2012)\citenamefont{Evans, Fiser, Prins,
  Rapp, Shields, Glass, and Superfine}}]{evans2012highly}
\bibinfo{author}{\bibfnamefont{B.~A.} \bibnamefont{Evans}},
  \bibinfo{author}{\bibfnamefont{B.~L.} \bibnamefont{Fiser}},
  \bibinfo{author}{\bibfnamefont{W.~J.} \bibnamefont{Prins}},
  \bibinfo{author}{\bibfnamefont{D.~J.} \bibnamefont{Rapp}},
  \bibinfo{author}{\bibfnamefont{A.~R.} \bibnamefont{Shields}},
  \bibinfo{author}{\bibfnamefont{D.~R.} \bibnamefont{Glass}}, \bibnamefont{and}
  \bibinfo{author}{\bibfnamefont{R.}~\bibnamefont{Superfine}},
  \bibinfo{journal}{J. Magn. Magn. Mater.} \textbf{\bibinfo{volume}{324}},
  \bibinfo{pages}{501} (\bibinfo{year}{2012}).

\bibitem[{\citenamefont{Han et~al.}(2013)\citenamefont{Han, Hong, and
  Faidley}}]{han2013field}
\bibinfo{author}{\bibfnamefont{Y.}~\bibnamefont{Han}},
  \bibinfo{author}{\bibfnamefont{W.}~\bibnamefont{Hong}}, \bibnamefont{and}
  \bibinfo{author}{\bibfnamefont{L.~E.} \bibnamefont{Faidley}},
  \bibinfo{journal}{Int. J. Solids Struct.} \textbf{\bibinfo{volume}{50}},
  \bibinfo{pages}{2281} (\bibinfo{year}{2013}).

\bibitem[{\citenamefont{Borin et~al.}(2013)\citenamefont{Borin, Stepanov, and
  Odenbach}}]{borin2013tuning}
\bibinfo{author}{\bibfnamefont{D.~Y.} \bibnamefont{Borin}},
  \bibinfo{author}{\bibfnamefont{G.~V.} \bibnamefont{Stepanov}},
  \bibnamefont{and} \bibinfo{author}{\bibfnamefont{S.}~\bibnamefont{Odenbach}},
  \bibinfo{journal}{J. Phys.: Conf. Ser.} \textbf{\bibinfo{volume}{412}},
  \bibinfo{pages}{012040} (\bibinfo{year}{2013}).

\bibitem[{\citenamefont{Chiba et~al.}(2013)\citenamefont{Chiba, Yamamoto, Hojo,
  Kawai, and Mitsumata}}]{chiba2013wide}
\bibinfo{author}{\bibfnamefont{N.}~\bibnamefont{Chiba}},
  \bibinfo{author}{\bibfnamefont{K.}~\bibnamefont{Yamamoto}},
  \bibinfo{author}{\bibfnamefont{T.}~\bibnamefont{Hojo}},
  \bibinfo{author}{\bibfnamefont{M.}~\bibnamefont{Kawai}}, \bibnamefont{and}
  \bibinfo{author}{\bibfnamefont{T.}~\bibnamefont{Mitsumata}},
  \bibinfo{journal}{Chem. Lett.} \textbf{\bibinfo{volume}{42}},
  \bibinfo{pages}{253} (\bibinfo{year}{2013}).

\bibitem[{\citenamefont{Pessot et~al.}(2014)\citenamefont{Pessot, Cremer,
  Borin, Odenbach, L{\"o}wen, and Menzel}}]{pessot2014structural}
\bibinfo{author}{\bibfnamefont{G.}~\bibnamefont{Pessot}},
  \bibinfo{author}{\bibfnamefont{P.}~\bibnamefont{Cremer}},
  \bibinfo{author}{\bibfnamefont{D.~Y.} \bibnamefont{Borin}},
  \bibinfo{author}{\bibfnamefont{S.}~\bibnamefont{Odenbach}},
  \bibinfo{author}{\bibfnamefont{H.}~\bibnamefont{L{\"o}wen}},
  \bibnamefont{and} \bibinfo{author}{\bibfnamefont{A.~M.}
  \bibnamefont{Menzel}}, \bibinfo{journal}{J. Chem. Phys.}
  \textbf{\bibinfo{volume}{141}}, \bibinfo{pages}{015005}
  (\bibinfo{year}{2014}).

\bibitem[{\citenamefont{Sorokin et~al.}(2015)\citenamefont{Sorokin, Stepanov,
  Shamonin, Monkman, Khokhlov, and Kramarenko}}]{sorokin2015hysteresis}
\bibinfo{author}{\bibfnamefont{V.~V.} \bibnamefont{Sorokin}},
  \bibinfo{author}{\bibfnamefont{G.~V.} \bibnamefont{Stepanov}},
  \bibinfo{author}{\bibfnamefont{M.}~\bibnamefont{Shamonin}},
  \bibinfo{author}{\bibfnamefont{G.~J.} \bibnamefont{Monkman}},
  \bibinfo{author}{\bibfnamefont{A.~R.} \bibnamefont{Khokhlov}},
  \bibnamefont{and} \bibinfo{author}{\bibfnamefont{E.~Y.}
  \bibnamefont{Kramarenko}}, \bibinfo{journal}{Polymer}
  \textbf{\bibinfo{volume}{76}}, \bibinfo{pages}{191} (\bibinfo{year}{2015}).

\bibitem[{\citenamefont{Pessot et~al.}(2016)\citenamefont{Pessot, L{\"o}wen,
  and Menzel}}]{pessot2016dynamic}
\bibinfo{author}{\bibfnamefont{G.}~\bibnamefont{Pessot}},
  \bibinfo{author}{\bibfnamefont{H.}~\bibnamefont{L{\"o}wen}},
  \bibnamefont{and} \bibinfo{author}{\bibfnamefont{A.~M.}
  \bibnamefont{Menzel}}, \bibinfo{journal}{J. Chem. Phys.}
  \textbf{\bibinfo{volume}{145}}, \bibinfo{pages}{104904}
  (\bibinfo{year}{2016}).

\bibitem[{\citenamefont{Volkova et~al.}(2017)\citenamefont{Volkova, B{\"o}hm,
  Kaufhold, Popp, Becker, Borin, Stepanov, and Zimmermann}}]{volkova2017motion}
\bibinfo{author}{\bibfnamefont{T.~I.} \bibnamefont{Volkova}},
  \bibinfo{author}{\bibfnamefont{V.}~\bibnamefont{B{\"o}hm}},
  \bibinfo{author}{\bibfnamefont{T.}~\bibnamefont{Kaufhold}},
  \bibinfo{author}{\bibfnamefont{J.}~\bibnamefont{Popp}},
  \bibinfo{author}{\bibfnamefont{F.}~\bibnamefont{Becker}},
  \bibinfo{author}{\bibfnamefont{D.~Y.} \bibnamefont{Borin}},
  \bibinfo{author}{\bibfnamefont{G.~V.} \bibnamefont{Stepanov}},
  \bibnamefont{and}
  \bibinfo{author}{\bibfnamefont{K.}~\bibnamefont{Zimmermann}},
  \bibinfo{journal}{J. Magn. Magn. Mater.} \textbf{\bibinfo{volume}{431}},
  \bibinfo{pages}{262} (\bibinfo{year}{2017}).

\bibitem[{\citenamefont{Cremer et~al.}(2015)\citenamefont{Cremer, L{\"o}wen,
  and Menzel}}]{cremer2015tailoring}
\bibinfo{author}{\bibfnamefont{P.}~\bibnamefont{Cremer}},
  \bibinfo{author}{\bibfnamefont{H.}~\bibnamefont{L{\"o}wen}},
  \bibnamefont{and} \bibinfo{author}{\bibfnamefont{A.~M.}
  \bibnamefont{Menzel}}, \bibinfo{journal}{Appl. Phys. Lett.}
  \textbf{\bibinfo{volume}{107}}, \bibinfo{pages}{171903}
  (\bibinfo{year}{2015}).

\bibitem[{\citenamefont{Cremer et~al.}(2016)\citenamefont{Cremer, L{\"o}wen,
  and Menzel}}]{cremer2016superelastic}
\bibinfo{author}{\bibfnamefont{P.}~\bibnamefont{Cremer}},
  \bibinfo{author}{\bibfnamefont{H.}~\bibnamefont{L{\"o}wen}},
  \bibnamefont{and} \bibinfo{author}{\bibfnamefont{A.~M.}
  \bibnamefont{Menzel}}, \bibinfo{journal}{Phys. Chem. Chem. Phys.}
  \textbf{\bibinfo{volume}{18}}, \bibinfo{pages}{26670} (\bibinfo{year}{2016}).

\bibitem[{\citenamefont{Annunziata et~al.}(2013)\citenamefont{Annunziata,
  Menzel, and L{\"o}wen}}]{annunziata2013hardening}
\bibinfo{author}{\bibfnamefont{M.~A.} \bibnamefont{Annunziata}},
  \bibinfo{author}{\bibfnamefont{A.~M.} \bibnamefont{Menzel}},
  \bibnamefont{and}
  \bibinfo{author}{\bibfnamefont{H.}~\bibnamefont{L{\"o}wen}},
  \bibinfo{journal}{J. Chem. Phys.} \textbf{\bibinfo{volume}{138}},
  \bibinfo{pages}{204906} (\bibinfo{year}{2013}).

\bibitem[{\citenamefont{Melenev et~al.}(2006)\citenamefont{Melenev, Rusakov,
  and Raikher}}]{melenev2006magnetic}
\bibinfo{author}{\bibfnamefont{P.~V.} \bibnamefont{Melenev}},
  \bibinfo{author}{\bibfnamefont{V.~V.} \bibnamefont{Rusakov}},
  \bibnamefont{and} \bibinfo{author}{\bibfnamefont{Y.~L.}
  \bibnamefont{Raikher}}, \bibinfo{journal}{J. Magn. Magn. Mater.}
  \textbf{\bibinfo{volume}{300}}, \bibinfo{pages}{e187} (\bibinfo{year}{2006}).

\bibitem[{\citenamefont{Stepanov et~al.}(2008)\citenamefont{Stepanov, Borin,
  Raikher, Melenev, and Perov}}]{stepanov2008motion}
\bibinfo{author}{\bibfnamefont{G.~V.} \bibnamefont{Stepanov}},
  \bibinfo{author}{\bibfnamefont{D.~Y.} \bibnamefont{Borin}},
  \bibinfo{author}{\bibfnamefont{Y.~L.} \bibnamefont{Raikher}},
  \bibinfo{author}{\bibfnamefont{P.~V.} \bibnamefont{Melenev}},
  \bibnamefont{and} \bibinfo{author}{\bibfnamefont{N.~S.} \bibnamefont{Perov}},
  \bibinfo{journal}{J. Phys.: Condens. Matter} \textbf{\bibinfo{volume}{20}},
  \bibinfo{pages}{204121} (\bibinfo{year}{2008}).

\bibitem[{\citenamefont{Biller et~al.}(2014)\citenamefont{Biller, Stolbov, and
  Raikher}}]{biller2014modeling}
\bibinfo{author}{\bibfnamefont{A.~M.} \bibnamefont{Biller}},
  \bibinfo{author}{\bibfnamefont{O.~V.} \bibnamefont{Stolbov}},
  \bibnamefont{and} \bibinfo{author}{\bibfnamefont{Y.~L.}
  \bibnamefont{Raikher}}, \bibinfo{journal}{J. Appl. Phys.}
  \textbf{\bibinfo{volume}{116}}, \bibinfo{pages}{114904}
  (\bibinfo{year}{2014}).

\bibitem[{\citenamefont{Biller et~al.}(2015)\citenamefont{Biller, Stolbov, and
  Raikher}}]{biller2015mesoscopic}
\bibinfo{author}{\bibfnamefont{A.~M.} \bibnamefont{Biller}},
  \bibinfo{author}{\bibfnamefont{O.~V.} \bibnamefont{Stolbov}},
  \bibnamefont{and} \bibinfo{author}{\bibfnamefont{Y.~L.}
  \bibnamefont{Raikher}}, \bibinfo{journal}{Phys. Rev. E}
  \textbf{\bibinfo{volume}{92}}, \bibinfo{pages}{023202}
  (\bibinfo{year}{2015}).

\bibitem[{\citenamefont{Zubarev et~al.}(2017)\citenamefont{Zubarev, Chirikov,
  Stepanov, and Borin}}]{zubarev2017hysteresis}
\bibinfo{author}{\bibfnamefont{A.}~\bibnamefont{Zubarev}},
  \bibinfo{author}{\bibfnamefont{D.}~\bibnamefont{Chirikov}},
  \bibinfo{author}{\bibfnamefont{G.}~\bibnamefont{Stepanov}}, \bibnamefont{and}
  \bibinfo{author}{\bibfnamefont{D.}~\bibnamefont{Borin}}, \bibinfo{journal}{J.
  Magn. Magn. Mater.} \textbf{\bibinfo{volume}{431}}, \bibinfo{pages}{120}
  (\bibinfo{year}{2017}).

\bibitem[{\citenamefont{Biller et~al.}(2018)\citenamefont{Biller, Stolbov, and
  Raikher}}]{biller2018two}
\bibinfo{author}{\bibfnamefont{A.~M.} \bibnamefont{Biller}},
  \bibinfo{author}{\bibfnamefont{O.~V.} \bibnamefont{Stolbov}},
  \bibnamefont{and} \bibinfo{author}{\bibfnamefont{Y.~L.}
  \bibnamefont{Raikher}}, \bibinfo{journal}{J. Phys.: Conf. Ser.}
  \textbf{\bibinfo{volume}{994}}, \bibinfo{pages}{012001}
  (\bibinfo{year}{2018}).

\bibitem[{\citenamefont{Crick and Hughes}(1950)}]{crick1950physical}
\bibinfo{author}{\bibfnamefont{F.~H.~C.} \bibnamefont{Crick}} \bibnamefont{and}
  \bibinfo{author}{\bibfnamefont{A.~F.~W.} \bibnamefont{Hughes}},
  \bibinfo{journal}{Exp. Cell Res.} \textbf{\bibinfo{volume}{1}},
  \bibinfo{pages}{37} (\bibinfo{year}{1950}).

\bibitem[{\citenamefont{Ziemann et~al.}(1994)\citenamefont{Ziemann, R{\"a}dler,
  and Sackmann}}]{ziemann1994local}
\bibinfo{author}{\bibfnamefont{F.}~\bibnamefont{Ziemann}},
  \bibinfo{author}{\bibfnamefont{J.}~\bibnamefont{R{\"a}dler}},
  \bibnamefont{and} \bibinfo{author}{\bibfnamefont{E.}~\bibnamefont{Sackmann}},
  \bibinfo{journal}{Biophys. J.} \textbf{\bibinfo{volume}{66}},
  \bibinfo{pages}{2210} (\bibinfo{year}{1994}).

\bibitem[{\citenamefont{Bausch et~al.}(1999)\citenamefont{Bausch, M{\"o}ller,
  and Sackmann}}]{bausch1999measurement}
\bibinfo{author}{\bibfnamefont{A.~R.} \bibnamefont{Bausch}},
  \bibinfo{author}{\bibfnamefont{W.}~\bibnamefont{M{\"o}ller}},
  \bibnamefont{and} \bibinfo{author}{\bibfnamefont{E.}~\bibnamefont{Sackmann}},
  \bibinfo{journal}{Biophys. J.} \textbf{\bibinfo{volume}{76}},
  \bibinfo{pages}{573} (\bibinfo{year}{1999}).

\bibitem[{\citenamefont{Waigh}(2005)}]{waigh2005microrheology}
\bibinfo{author}{\bibfnamefont{T.~A.} \bibnamefont{Waigh}},
  \bibinfo{journal}{Rep. Prog. Phys.} \textbf{\bibinfo{volume}{68}},
  \bibinfo{pages}{685} (\bibinfo{year}{2005}).

\bibitem[{\citenamefont{Wilhelm}(2008)}]{wilhelm2008out}
\bibinfo{author}{\bibfnamefont{C.}~\bibnamefont{Wilhelm}},
  \bibinfo{journal}{Phys. Rev. Lett.} \textbf{\bibinfo{volume}{101}},
  \bibinfo{pages}{028101} (\bibinfo{year}{2008}).

\bibitem[{\citenamefont{Chippada et~al.}(2009)\citenamefont{Chippada, Langrana,
  and Yurke}}]{chippada2009complete}
\bibinfo{author}{\bibfnamefont{U.}~\bibnamefont{Chippada}},
  \bibinfo{author}{\bibfnamefont{N.}~\bibnamefont{Langrana}}, \bibnamefont{and}
  \bibinfo{author}{\bibfnamefont{B.}~\bibnamefont{Yurke}}, \bibinfo{journal}{J.
  Appl. Phys.} \textbf{\bibinfo{volume}{106}}, \bibinfo{pages}{063528}
  (\bibinfo{year}{2009}).

\bibitem[{\citenamefont{Roeder et~al.}(2012)\citenamefont{Roeder, Bender,
  Tsch{\"o}pe, Birringer, and Schmidt}}]{roeder2012shear}
\bibinfo{author}{\bibfnamefont{L.}~\bibnamefont{Roeder}},
  \bibinfo{author}{\bibfnamefont{P.}~\bibnamefont{Bender}},
  \bibinfo{author}{\bibfnamefont{A.}~\bibnamefont{Tsch{\"o}pe}},
  \bibinfo{author}{\bibfnamefont{R.}~\bibnamefont{Birringer}},
  \bibnamefont{and} \bibinfo{author}{\bibfnamefont{A.~M.}
  \bibnamefont{Schmidt}}, \bibinfo{journal}{J. Polym. Sci. B: Polym. Phys.}
  \textbf{\bibinfo{volume}{50}}, \bibinfo{pages}{1772} (\bibinfo{year}{2012}).

\bibitem[{\citenamefont{Bender et~al.}(2013)\citenamefont{Bender, Tsch{\"o}pe,
  and Birringer}}]{bender2013determination}
\bibinfo{author}{\bibfnamefont{P.}~\bibnamefont{Bender}},
  \bibinfo{author}{\bibfnamefont{A.}~\bibnamefont{Tsch{\"o}pe}},
  \bibnamefont{and}
  \bibinfo{author}{\bibfnamefont{R.}~\bibnamefont{Birringer}},
  \bibinfo{journal}{J. Magn. Magn. Mater.} \textbf{\bibinfo{volume}{346}},
  \bibinfo{pages}{152} (\bibinfo{year}{2013}).

\bibitem[{\citenamefont{Huang et~al.}(2017)\citenamefont{Huang, Gawlitza, von
  Klitzing, Steffen, and Auernhammer}}]{huang2017structure}
\bibinfo{author}{\bibfnamefont{S.}~\bibnamefont{Huang}},
  \bibinfo{author}{\bibfnamefont{K.}~\bibnamefont{Gawlitza}},
  \bibinfo{author}{\bibfnamefont{R.}~\bibnamefont{von Klitzing}},
  \bibinfo{author}{\bibfnamefont{W.}~\bibnamefont{Steffen}}, \bibnamefont{and}
  \bibinfo{author}{\bibfnamefont{G.~K.} \bibnamefont{Auernhammer}},
  \bibinfo{journal}{Macromolecules} \textbf{\bibinfo{volume}{50}},
  \bibinfo{pages}{3680} (\bibinfo{year}{2017}).

\bibitem[{\citenamefont{An et~al.}(2014)\citenamefont{An, Groenewold, Picken,
  and Mendes}}]{an2014conformational}
\bibinfo{author}{\bibfnamefont{H.-N.} \bibnamefont{An}},
  \bibinfo{author}{\bibfnamefont{J.}~\bibnamefont{Groenewold}},
  \bibinfo{author}{\bibfnamefont{S.~J.} \bibnamefont{Picken}},
  \bibnamefont{and} \bibinfo{author}{\bibfnamefont{E.}~\bibnamefont{Mendes}},
  \bibinfo{journal}{Soft Matter} \textbf{\bibinfo{volume}{10}},
  \bibinfo{pages}{997} (\bibinfo{year}{2014}).

\bibitem[{\citenamefont{Gundermann and
  Odenbach}(2014)}]{gundermann2014investigation}
\bibinfo{author}{\bibfnamefont{T.}~\bibnamefont{Gundermann}} \bibnamefont{and}
  \bibinfo{author}{\bibfnamefont{S.}~\bibnamefont{Odenbach}},
  \bibinfo{journal}{Smart Mater. Struct.} \textbf{\bibinfo{volume}{23}},
  \bibinfo{pages}{105013} (\bibinfo{year}{2014}).

\bibitem[{\citenamefont{Gundermann et~al.}(2017)\citenamefont{Gundermann,
  Cremer, L\"owen, Menzel, and Odenbach}}]{gundermann2017statistical}
\bibinfo{author}{\bibfnamefont{T.}~\bibnamefont{Gundermann}},
  \bibinfo{author}{\bibfnamefont{P.}~\bibnamefont{Cremer}},
  \bibinfo{author}{\bibfnamefont{H.}~\bibnamefont{L\"owen}},
  \bibinfo{author}{\bibfnamefont{A.~M.} \bibnamefont{Menzel}},
  \bibnamefont{and} \bibinfo{author}{\bibfnamefont{S.}~\bibnamefont{Odenbach}},
  \bibinfo{journal}{Smart Mater. Struct.} \textbf{\bibinfo{volume}{26}},
  \bibinfo{pages}{045012} (\bibinfo{year}{2017}).

\bibitem[{\citenamefont{Sch{\"u}mann et~al.}(2017)\citenamefont{Sch{\"u}mann,
  Borin, Huang, Auernhammer, M{\"u}ller, and
  Odenbach}}]{schumann2017characterization}
\bibinfo{author}{\bibfnamefont{M.}~\bibnamefont{Sch{\"u}mann}},
  \bibinfo{author}{\bibfnamefont{D.}~\bibnamefont{Borin}},
  \bibinfo{author}{\bibfnamefont{S.}~\bibnamefont{Huang}},
  \bibinfo{author}{\bibfnamefont{G.}~\bibnamefont{Auernhammer}},
  \bibinfo{author}{\bibfnamefont{R.}~\bibnamefont{M{\"u}ller}},
  \bibnamefont{and} \bibinfo{author}{\bibfnamefont{S.}~\bibnamefont{Odenbach}},
  \bibinfo{journal}{Smart Mater. Struct.} \textbf{\bibinfo{volume}{26}},
  \bibinfo{pages}{095018} (\bibinfo{year}{2017}).

\bibitem[{\citenamefont{Tarama et~al.}(2014)\citenamefont{Tarama, Cremer,
  Borin, Odenbach, L{\"o}wen, and Menzel}}]{tarama2014tunable}
\bibinfo{author}{\bibfnamefont{M.}~\bibnamefont{Tarama}},
  \bibinfo{author}{\bibfnamefont{P.}~\bibnamefont{Cremer}},
  \bibinfo{author}{\bibfnamefont{D.~Y.} \bibnamefont{Borin}},
  \bibinfo{author}{\bibfnamefont{S.}~\bibnamefont{Odenbach}},
  \bibinfo{author}{\bibfnamefont{H.}~\bibnamefont{L{\"o}wen}},
  \bibnamefont{and} \bibinfo{author}{\bibfnamefont{A.~M.}
  \bibnamefont{Menzel}}, \bibinfo{journal}{Phys. Rev. E}
  \textbf{\bibinfo{volume}{90}}, \bibinfo{pages}{042311}
  (\bibinfo{year}{2014}).

\bibitem[{\citenamefont{Ivaneyko et~al.}(2015)\citenamefont{Ivaneyko,
  Toshchevikov, and Saphiannikova}}]{ivaneyko2015dynamic}
\bibinfo{author}{\bibfnamefont{D.}~\bibnamefont{Ivaneyko}},
  \bibinfo{author}{\bibfnamefont{V.}~\bibnamefont{Toshchevikov}},
  \bibnamefont{and}
  \bibinfo{author}{\bibfnamefont{M.}~\bibnamefont{Saphiannikova}},
  \bibinfo{journal}{Soft Matter} \textbf{\bibinfo{volume}{11}},
  \bibinfo{pages}{7627} (\bibinfo{year}{2015}).

\bibitem[{\citenamefont{Cremer et~al.}(2017)\citenamefont{Cremer, Heinen,
  Menzel, and L{\"o}wen}}]{cremer2017density}
\bibinfo{author}{\bibfnamefont{P.}~\bibnamefont{Cremer}},
  \bibinfo{author}{\bibfnamefont{M.}~\bibnamefont{Heinen}},
  \bibinfo{author}{\bibfnamefont{A.~M.} \bibnamefont{Menzel}},
  \bibnamefont{and}
  \bibinfo{author}{\bibfnamefont{H.}~\bibnamefont{L{\"o}wen}},
  \bibinfo{journal}{J. Phys.: Condens. Matter} \textbf{\bibinfo{volume}{29}},
  \bibinfo{pages}{275102} (\bibinfo{year}{2017}).

\bibitem[{\citenamefont{Pessot et~al.}(2018)\citenamefont{Pessot, Sch{\"u}mann,
  Gundermann, Odenbach, L{\"o}wen, and Menzel}}]{pessot2018tunable}
\bibinfo{author}{\bibfnamefont{G.}~\bibnamefont{Pessot}},
  \bibinfo{author}{\bibfnamefont{M.}~\bibnamefont{Sch{\"u}mann}},
  \bibinfo{author}{\bibfnamefont{T.}~\bibnamefont{Gundermann}},
  \bibinfo{author}{\bibfnamefont{S.}~\bibnamefont{Odenbach}},
  \bibinfo{author}{\bibfnamefont{H.}~\bibnamefont{L{\"o}wen}},
  \bibnamefont{and} \bibinfo{author}{\bibfnamefont{A.~M.}
  \bibnamefont{Menzel}}, \bibinfo{journal}{J. Phys.: Condens. Matter}
  \textbf{\bibinfo{volume}{30}}, \bibinfo{pages}{125101}
  (\bibinfo{year}{2018}).

\bibitem[{\citenamefont{Kalina et~al.}(2016)\citenamefont{Kalina, Metsch, and
  K{\"a}stner}}]{kalina2016microscale}
\bibinfo{author}{\bibfnamefont{K.~A.} \bibnamefont{Kalina}},
  \bibinfo{author}{\bibfnamefont{P.}~\bibnamefont{Metsch}}, \bibnamefont{and}
  \bibinfo{author}{\bibfnamefont{M.}~\bibnamefont{K{\"a}stner}},
  \bibinfo{journal}{Int. J. Solids Struct.} \textbf{\bibinfo{volume}{102}},
  \bibinfo{pages}{286} (\bibinfo{year}{2016}).

\bibitem[{\citenamefont{Attaran et~al.}(2017)\citenamefont{Attaran, Brummund,
  and Wallmersperger}}]{attaran2017modeling}
\bibinfo{author}{\bibfnamefont{A.}~\bibnamefont{Attaran}},
  \bibinfo{author}{\bibfnamefont{J.}~\bibnamefont{Brummund}}, \bibnamefont{and}
  \bibinfo{author}{\bibfnamefont{T.}~\bibnamefont{Wallmersperger}},
  \bibinfo{journal}{J. Magn. Magn. Mater.} \textbf{\bibinfo{volume}{431}},
  \bibinfo{pages}{188} (\bibinfo{year}{2017}).

\bibitem[{\citenamefont{Phan-Thien}(1993)}]{phan1993rigid}
\bibinfo{author}{\bibfnamefont{N.}~\bibnamefont{Phan-Thien}},
  \bibinfo{journal}{J. Elasticity} \textbf{\bibinfo{volume}{32}},
  \bibinfo{pages}{243} (\bibinfo{year}{1993}).

\bibitem[{\citenamefont{Kim and Phan-Thien}(1995)}]{kim1994faxen}
\bibinfo{author}{\bibfnamefont{S.}~\bibnamefont{Kim}} \bibnamefont{and}
  \bibinfo{author}{\bibfnamefont{N.}~\bibnamefont{Phan-Thien}},
  \bibinfo{journal}{J. Elasticity} \textbf{\bibinfo{volume}{37}},
  \bibinfo{pages}{93} (\bibinfo{year}{1995}).

\bibitem[{\citenamefont{Phan-Thien and Kim}(1994)}]{phan1994load}
\bibinfo{author}{\bibfnamefont{N.}~\bibnamefont{Phan-Thien}} \bibnamefont{and}
  \bibinfo{author}{\bibfnamefont{S.}~\bibnamefont{Kim}},
  \bibinfo{journal}{ZAMP} \textbf{\bibinfo{volume}{45}}, \bibinfo{pages}{177}
  (\bibinfo{year}{1994}).

\bibitem[{\citenamefont{Norris}(2008)}]{norris2008faxen}
\bibinfo{author}{\bibfnamefont{A.~N.} \bibnamefont{Norris}},
  \bibinfo{journal}{J. Acoust. Soc. Am.} \textbf{\bibinfo{volume}{123}},
  \bibinfo{pages}{99} (\bibinfo{year}{2008}).

\bibitem[{\citenamefont{Puljiz et~al.}(2016)\citenamefont{Puljiz, Huang,
  Auernhammer, and Menzel}}]{puljiz2016forces}
\bibinfo{author}{\bibfnamefont{M.}~\bibnamefont{Puljiz}},
  \bibinfo{author}{\bibfnamefont{S.}~\bibnamefont{Huang}},
  \bibinfo{author}{\bibfnamefont{G.~K.} \bibnamefont{Auernhammer}},
  \bibnamefont{and} \bibinfo{author}{\bibfnamefont{A.~M.}
  \bibnamefont{Menzel}}, \bibinfo{journal}{Phys. Rev. Lett.}
  \textbf{\bibinfo{volume}{117}}, \bibinfo{pages}{238003}
  (\bibinfo{year}{2016}).

\bibitem[{\citenamefont{Puljiz and Menzel}(2017)}]{puljiz2017forces}
\bibinfo{author}{\bibfnamefont{M.}~\bibnamefont{Puljiz}} \bibnamefont{and}
  \bibinfo{author}{\bibfnamefont{A.~M.} \bibnamefont{Menzel}},
  \bibinfo{journal}{Phys. Rev. E} \textbf{\bibinfo{volume}{95}},
  \bibinfo{pages}{053002} (\bibinfo{year}{2017}).

\bibitem[{\citenamefont{Menzel}(2017)}]{menzel2017force}
\bibinfo{author}{\bibfnamefont{A.~M.} \bibnamefont{Menzel}},
  \bibinfo{journal}{Soft Matter} \textbf{\bibinfo{volume}{13}},
  \bibinfo{pages}{3373} (\bibinfo{year}{2017}).

\bibitem[{\citenamefont{Klapp}(2005)}]{klapp2005dipolar}
\bibinfo{author}{\bibfnamefont{S.~H.~L.} \bibnamefont{Klapp}},
  \bibinfo{journal}{J. Phys.: Condens. Matter} \textbf{\bibinfo{volume}{17}},
  \bibinfo{pages}{R525} (\bibinfo{year}{2005}).

\bibitem[{\citenamefont{Holm and Weis}(2005)}]{holm2005structure}
\bibinfo{author}{\bibfnamefont{C.}~\bibnamefont{Holm}} \bibnamefont{and}
  \bibinfo{author}{\bibfnamefont{J.-J.} \bibnamefont{Weis}},
  \bibinfo{journal}{Curr. Opin. Colloid Interf. Sci.}
  \textbf{\bibinfo{volume}{10}}, \bibinfo{pages}{133} (\bibinfo{year}{2005}).

\bibitem[{\citenamefont{Menzel}(2014)}]{menzel2014bridging}
\bibinfo{author}{\bibfnamefont{A.~M.} \bibnamefont{Menzel}},
  \bibinfo{journal}{J. Chem. Phys.} \textbf{\bibinfo{volume}{141}},
  \bibinfo{pages}{194907} (\bibinfo{year}{2014}).

\bibitem[{\citenamefont{Pessot et~al.}(2015)\citenamefont{Pessot, Weeber, Holm,
  L{\"o}wen, and Menzel}}]{pessot2015towards}
\bibinfo{author}{\bibfnamefont{G.}~\bibnamefont{Pessot}},
  \bibinfo{author}{\bibfnamefont{R.}~\bibnamefont{Weeber}},
  \bibinfo{author}{\bibfnamefont{C.}~\bibnamefont{Holm}},
  \bibinfo{author}{\bibfnamefont{H.}~\bibnamefont{L{\"o}wen}},
  \bibnamefont{and} \bibinfo{author}{\bibfnamefont{A.~M.}
  \bibnamefont{Menzel}}, \bibinfo{journal}{J. Phys.: Condens. Matter}
  \textbf{\bibinfo{volume}{27}}, \bibinfo{pages}{325105}
  (\bibinfo{year}{2015}).

\bibitem[{\citenamefont{Zubarev}(2013{\natexlab{b}})}]{zubarev2013magnetodeformation}
\bibinfo{author}{\bibfnamefont{A.}~\bibnamefont{Zubarev}},
  \bibinfo{journal}{Physica A} \textbf{\bibinfo{volume}{392}},
  \bibinfo{pages}{4824} (\bibinfo{year}{2013}{\natexlab{b}}).

\bibitem[{\citenamefont{Romeis et~al.}(2017)\citenamefont{Romeis, Metsch,
  K{\"a}stner, and Saphiannikova}}]{romeis2017theoretical}
\bibinfo{author}{\bibfnamefont{D.}~\bibnamefont{Romeis}},
  \bibinfo{author}{\bibfnamefont{P.}~\bibnamefont{Metsch}},
  \bibinfo{author}{\bibfnamefont{M.}~\bibnamefont{K{\"a}stner}},
  \bibnamefont{and}
  \bibinfo{author}{\bibfnamefont{M.}~\bibnamefont{Saphiannikova}},
  \bibinfo{journal}{Phys. Rev. E} \textbf{\bibinfo{volume}{95}},
  \bibinfo{pages}{042501} (\bibinfo{year}{2017}).

\bibitem[{ima()}]{imagej}
\bibinfo{note}{\lowercase{h}ttp://imagej.nih.gov/ij/}.

\bibitem[{\citenamefont{Landau and Lifshitz}(1986)}]{landau1986theory}
\bibinfo{author}{\bibfnamefont{L.~D.} \bibnamefont{Landau}} \bibnamefont{and}
  \bibinfo{author}{\bibfnamefont{E.~M.} \bibnamefont{Lifshitz}},
  \emph{\bibinfo{title}{Theory of {Elasticity}}} (\bibinfo{publisher}{Elsevier,
  Oxford}, \bibinfo{year}{1986}).

\bibitem[{\citenamefont{Press et~al.}(1982)\citenamefont{Press, Teukolsky,
  Vetterling, and Flannery}}]{press1982numerical}
\bibinfo{author}{\bibfnamefont{W.~H.} \bibnamefont{Press}},
  \bibinfo{author}{\bibfnamefont{S.~A.} \bibnamefont{Teukolsky}},
  \bibinfo{author}{\bibfnamefont{W.~T.} \bibnamefont{Vetterling}},
  \bibnamefont{and} \bibinfo{author}{\bibfnamefont{B.~P.}
  \bibnamefont{Flannery}}, \emph{\bibinfo{title}{Numerical Recipes in C}}
  (\bibinfo{publisher}{Cambridge University Press, Cambridge},
  \bibinfo{year}{1982}).

\bibitem[{\citenamefont{Kalina et~al.}(2017)\citenamefont{Kalina, Brummund,
  Metsch, K{\"a}stner, Borin, Linke, and Odenbach}}]{Kalina2017}
\bibinfo{author}{\bibfnamefont{K.~A.} \bibnamefont{Kalina}},
  \bibinfo{author}{\bibfnamefont{J.}~\bibnamefont{Brummund}},
  \bibinfo{author}{\bibfnamefont{P.}~\bibnamefont{Metsch}},
  \bibinfo{author}{\bibfnamefont{M.}~\bibnamefont{K{\"a}stner}},
  \bibinfo{author}{\bibfnamefont{D.~Y.} \bibnamefont{Borin}},
  \bibinfo{author}{\bibfnamefont{J.~M.} \bibnamefont{Linke}}, \bibnamefont{and}
  \bibinfo{author}{\bibfnamefont{S.}~\bibnamefont{Odenbach}},
  \bibinfo{journal}{Smart Mater. Struct.} \textbf{\bibinfo{volume}{26}},
  \bibinfo{pages}{105019} (\bibinfo{year}{2017}).

\bibitem[{\citenamefont{de~Groot and Suttorp}(1972)}]{Groot1972}
\bibinfo{author}{\bibfnamefont{S.~R.} \bibnamefont{de~Groot}} \bibnamefont{and}
  \bibinfo{author}{\bibfnamefont{L.~G.} \bibnamefont{Suttorp}},
  \emph{\bibinfo{title}{Foundations of Electrodynamics}}
  (\bibinfo{publisher}{North-Holland}, \bibinfo{address}{Amsterdam},
  \bibinfo{year}{1972}).

\bibitem[{\citenamefont{Eringen and Maugin}(1990)}]{Eringen1990}
\bibinfo{author}{\bibfnamefont{A.~C.} \bibnamefont{Eringen}} \bibnamefont{and}
  \bibinfo{author}{\bibfnamefont{G.~A.} \bibnamefont{Maugin}},
  \emph{\bibinfo{title}{Electrodynamics of Continua I: Foundations and Solid
  Media}} (\bibinfo{publisher}{Springer}, \bibinfo{address}{New York},
  \bibinfo{year}{1990}).

\bibitem[{\citenamefont{Kankanala and Triantafyllidis}(2004)}]{Kankanala2004}
\bibinfo{author}{\bibfnamefont{S.~V.} \bibnamefont{Kankanala}}
  \bibnamefont{and}
  \bibinfo{author}{\bibfnamefont{N.}~\bibnamefont{Triantafyllidis}},
  \bibinfo{journal}{J. Mech. Phys. Solids} \textbf{\bibinfo{volume}{52}},
  \bibinfo{pages}{2869} (\bibinfo{year}{2004}).

\bibitem[{\citenamefont{Ogden}(2011)}]{Ogden2011book}
\bibinfo{author}{\bibfnamefont{R.~W.} \bibnamefont{Ogden}},
  \emph{\bibinfo{title}{Mechanics and Electrodynamics of Magneto- and
  Electro-Elastic Materials}} (\bibinfo{publisher}{Springer, Wien / New York},
  \bibinfo{year}{2011}).

\bibitem[{\citenamefont{Ponte~Casta{\~n}eda and
  Galipeau}(2011)}]{PonteCastaneda2011}
\bibinfo{author}{\bibfnamefont{P.}~\bibnamefont{Ponte~Casta{\~n}eda}}
  \bibnamefont{and} \bibinfo{author}{\bibfnamefont{E.}~\bibnamefont{Galipeau}},
  \bibinfo{journal}{J. Mech. Phys. Solids} \textbf{\bibinfo{volume}{59}},
  \bibinfo{pages}{194} (\bibinfo{year}{2011}).

\bibitem[{\citenamefont{Javili et~al.}(2013)\citenamefont{Javili,
  Chatzigeorgiou, and Steinmann}}]{Javili2013}
\bibinfo{author}{\bibfnamefont{A.}~\bibnamefont{Javili}},
  \bibinfo{author}{\bibfnamefont{G.}~\bibnamefont{Chatzigeorgiou}},
  \bibnamefont{and}
  \bibinfo{author}{\bibfnamefont{P.}~\bibnamefont{Steinmann}},
  \bibinfo{journal}{Int. J. Solids Struct.} \textbf{\bibinfo{volume}{50}},
  \bibinfo{pages}{4197} (\bibinfo{year}{2013}).

\bibitem[{\citenamefont{Danas}(2017)}]{Danas2017}
\bibinfo{author}{\bibfnamefont{K.}~\bibnamefont{Danas}}, \bibinfo{journal}{J.
  Mech. Phys. Solids} \textbf{\bibinfo{volume}{105}}, \bibinfo{pages}{25 }
  (\bibinfo{year}{2017}).

\bibitem[{\citenamefont{Grote and Feldhusen}(2011)}]{Dubbel2011}
\bibinfo{author}{\bibfnamefont{K.}~\bibnamefont{Grote}} \bibnamefont{and}
  \bibinfo{author}{\bibfnamefont{J.}~\bibnamefont{Feldhusen}},
  \emph{\bibinfo{title}{Dubbel: Taschenbuch f{\"u}r den Maschinenbau}}
  (\bibinfo{publisher}{Springer, Berlin / Heidelberg}, \bibinfo{year}{2011}).

\bibitem[{\citenamefont{Schwarz and Safran}(2002)}]{schwarz2002elastic}
\bibinfo{author}{\bibfnamefont{U.~S.} \bibnamefont{Schwarz}} \bibnamefont{and}
  \bibinfo{author}{\bibfnamefont{S.~A.} \bibnamefont{Safran}},
  \bibinfo{journal}{Phys. Rev. Lett.} \textbf{\bibinfo{volume}{88}},
  \bibinfo{pages}{048102} (\bibinfo{year}{2002}).

\bibitem[{\citenamefont{Schwarz and Safran}(2013)}]{schwarz2013physics}
\bibinfo{author}{\bibfnamefont{U.~S.} \bibnamefont{Schwarz}} \bibnamefont{and}
  \bibinfo{author}{\bibfnamefont{S.~A.} \bibnamefont{Safran}},
  \bibinfo{journal}{Rev. Mod. Phys.} \textbf{\bibinfo{volume}{85}},
  \bibinfo{pages}{1327} (\bibinfo{year}{2013}).

\bibitem[{\citenamefont{Bischofs et~al.}(2004)\citenamefont{Bischofs, Safran,
  and Schwarz}}]{bischofs2004elastic}
\bibinfo{author}{\bibfnamefont{I.~B.} \bibnamefont{Bischofs}},
  \bibinfo{author}{\bibfnamefont{S.~A.} \bibnamefont{Safran}},
  \bibnamefont{and} \bibinfo{author}{\bibfnamefont{U.~S.}
  \bibnamefont{Schwarz}}, \bibinfo{journal}{Phys. Rev. E}
  \textbf{\bibinfo{volume}{69}}, \bibinfo{pages}{021911}
  (\bibinfo{year}{2004}).

\bibitem[{\citenamefont{Yuval and Safran}(2013)}]{yuval2013dynamics}
\bibinfo{author}{\bibfnamefont{J.}~\bibnamefont{Yuval}} \bibnamefont{and}
  \bibinfo{author}{\bibfnamefont{S.~A.} \bibnamefont{Safran}},
  \bibinfo{journal}{Phys. Rev. E} \textbf{\bibinfo{volume}{87}},
  \bibinfo{pages}{042703} (\bibinfo{year}{2013}).

\bibitem[{\citenamefont{Teodosiu}(1982)}]{teodosiu1982elastic}
\bibinfo{author}{\bibfnamefont{C.}~\bibnamefont{Teodosiu}},
  \emph{\bibinfo{title}{The Elastic Field of Point Defects}}
  (\bibinfo{publisher}{Springer Berlin/Heidelberg}, \bibinfo{year}{1982}).

\bibitem[{\citenamefont{Boltz and Klumpp}(2017)}]{boltz2017buckling}
\bibinfo{author}{\bibfnamefont{H.-H.} \bibnamefont{Boltz}} \bibnamefont{and}
  \bibinfo{author}{\bibfnamefont{S.}~\bibnamefont{Klumpp}},
  \bibinfo{journal}{Eur. Phys. J. E} \textbf{\bibinfo{volume}{40}},
  \bibinfo{pages}{86} (\bibinfo{year}{2017}).

\end{thebibliography}

\end{document}